# Electron-phonon coupling in lattice engineering of lithium niobate single crystal thin films

Guoqiang Shi[1, 4, 6], Kunfeng Chen[2, 6], Hui Hu[3], Gongbin Tang[2], Dongfeng Xue[1, 5, ✉]

[1]Multiscale Crystal Materials Research Center, Shenzhen Institute of Advanced Technology, Chinese Academy of Sciences, Shenzhen 518055, China
[2]Institute of Novel Semiconductors, State Key Laboratory of Crystal Materials, Shandong University, Jinan 250100, China
[3]School of Physics, State Key Laboratory of Crystal Materials, Shandong University, Jinan 250100, China
[4] Shenzhen Key Laboratory of New Information Display and Storage Materials, College of Materials Science and Engineering, Shenzhen University, Shenzhen 518060, China
[5] Shenzhen Institute for Advanced Study, University of Electronic Science and Technology of China, Shenzhen 518110, China
[6]These authors contributed equally: Guoqiang Shi, Kunfeng Chen
✉e-mail: dfxue@uestc.edu.cn

**Abstract:**

Lithium niobate (LN) single crystal thin films are a high-performance photonic platform with applications in electro-optic modulators, nonlinear optical devices, optical frequency combs, and acousto-optic modulators. LN's significance in photonics parallels silicon's in electronics, addressing challenges like high power consumption and slow communication speeds, and offering potential for broad applications in optical communications, quantum computing, and artificial intelligence. Despite progress in developing LN-based photonic structures, achieving low-loss, reconfigurable, and large-scale devices requires improved processing techniques. This work introduces a quantum design methodology based on LN's crystal structure, utilizing electron-phonon coupling through external field perturbations. Multiscale structural analysis is performed with techniques such as time-of-flight secondary ion mass spectrometry, aberration-corrected transmission electron microscopy, and X-ray absorption spectra to identify and control defect structures. Angle-resolved Raman spectroscopy, femtosecond transient absorption spectroscopy, and Density Functional Theory further reveal the mechanisms of electron-phonon coupling. These findings establish a framework for designing LN-based quantum devices with enhanced performance and diverse functionalities.

**Introduction:**

Lithium niobate (LN) single crystal thin films, with its scale and density of photonic integration approaching that of semiconductor platforms, has emerged as a highly promising platform for integrated photonics[1-7]. Due to the material properties of LN, important active and passive photonic functions can be uniformly integrated in LN single crystal thin films without the need for additional materials[1,7-13]. Furthermore, the flexibility in designing waveguide geometry and dispersion in LN single crystal thin films allows for greater customization of components within integrated systems[8,14,15]. Consequently, LN single crystal thin films offer significant advantages in various applications, including electro-optic modulators[3,16-18], efficient nonlinear optical devices[7,12,17,19-22], optical frequency combs[5,9,22-25], and acousto-optic modulators[26]. Comparable in significance to silicon in the field of electronics, LN has the potential to replace silicon materials in photonics, *i.e.* "*Now entering, Lithium Niobate Valley*"[27], in which outstanding challenges in communication, such as high power and low speed, can be effectively addressed. Notably, quantum devices based on LN single crystal, including LN quantum light sources, LN quantum relays, and LN single-photon detectors, have become important trends in LN single crystal thin film applications[13]. The unique characteristics of LN single crystal thin films, including extremely low optical loss and abundant optoelectronic functionalities, make LN photonics chips highly attractive candidates in various fields, such as optical communications, data centers, photonic quantum computing, quantum communication, and artificial intelligence[14,21].

In recent years, significant progress has been made in the development of chip-scale micro-nano photonics structures based on LN single crystal thin films, thanks to the maturity of micro-nano fabrication processes and the availability of LN single crystal thin film platforms[4,17,26,28-31]. However, the performance enhancement of LN single crystal thin films as photonics platforms still relies on device fabrication processes, particularly in achieving low loss, reconfigurability, and large-scale devices. However, it is worth considering whether it is possible to construct LN-based devices by manipulating the quantum origins of LN single crystal thin films through the modulation of the LN crystal structure, thus bypassing the limitations imposed by device fabrication processes. Multiple degrees of freedom quantum design based on the coupling and decoupling processes among key degrees of freedom in crystal materials, including lattice, charge, spin, and orbital, offers a promising approach for constructing photonics platform devices with novel magnetic, electrical, optical, and other properties[32,33]. Based on this concept, we propose a novel quantum design approach that induces the coupling of multiple degrees of freedom in LN single crystal thin films through external field perturbation, enabling the manipulation of electron-phonon coupling in LN single crystal thin films using lasers for constructing photonic quantum devices.

Here we first performed comprehensive multiscale analysis of the structure of LN single crystal thin films. Advanced techniques such as time-of-flight secondary ion mass spectrometry (TOF-SIMS), aberration-corrected transmission electron microscopy (ACTEM), and X-ray absorption spectroscopy (XAS) were employed to elucidate the types and concentrations of defects. Utilizing defect structure lattice engineering, we further induced and characterized the electron-phonon coupling in LN single crystal thin films using angle-resolved Raman spectroscopy and femtosecond transient absorption spectroscopy (fs-TAS). Additionally, density functional theory (DFT) combined with defect structure models was employed to analyze the mechanisms controlling the electron-phonon coupling in LN single crystal thin films. Our research results demonstrate the effectiveness of combining multiscale structure analysis and design with the realization of electron-phonon coupling in LN single crystal thin films through external field perturbation, laying a solid foundation for efficient strategies in achieving multiple degrees of freedom quantum design. By directly exploring the quantum origins of LN single crystal thin film properties, our approach establishes a solid foundation for the design of controllable coupling LN quantum devices. This study opens up new pathways for the development of high-performance photonics devices and paves the way for advanced LN-based technologies with enhanced functionalities.

## Results and discussion

**Defects in LN single crystal thin films.** LN is a well-known material for its excellent properties such as high electro-optic, piezoelectric, and nonlinear optical coefficients[34]. However, its performance can be affected by defects that can arise during the crystal growth process[35]. One common type of defect in LN is known as the $Nb_{Li}$ antisite defect. The $Nb_{Li}$ antisite defect occurs when a Nb ion is substituted for a Li ion in the crystal lattice[34]. This defect creates an electrically active site, which can lead to electronic and optical performance degradation in the crystal. In addition, the presence of $Nb_{Li}$ antisite defects can increase the concentration of point defects, such as oxygen vacancies, in the crystal. Researchers have been studying the $Nb_{Li}$ antisite defect in LN crystals and its impact on the material's performance. Various techniques, such as electron paramagnetic resonance spectroscopy, optical absorption spectroscopy, and high-resolution transmission electron microscopy, have been used to characterize the $Nb_{Li}$ antisite defect and its effect on the crystal's properties[34]. To mitigate the impact of $Nb_{Li}$ antisite defects, researchers have investigated various approaches, including optimizing the crystal growth conditions, controlling the doping concentration, and annealing the crystal to reduce the defect concentration[36]. These methods have shown promising results in reducing the impact of $Nb_{Li}$ antisite defects on the crystal's performance.

The multilevel structure of the LN single crystal thin film was further examined by TOF-SIMS depth analysis, which was established by etching a selected 50μm×50μm area with continuous Ar-ion sputtering. Fig. 1a shows the TOF-SIMS depth profiles of various chemical species as a function of the sputtering time. The Li ion signal increases sharply during the first 30 seconds, and it remained stable between 30 seconds and 1169 seconds, then began to decline sharply. The intensity of the Nb ion increases in the first 17 seconds, and it remained stable between 17 seconds and 1231 seconds, then began to decline sharply. In addition, the presence of both $Li_2O$ and NbO is observed. The above phenomena indicate that Li and Nb exists as $Li_2O$ and NbO-based compounds, respectively. The ion mapping images of the Li and $Li_2O$ fragment (Fig. 1b) show that Li diffuses to the crystal surface after sputtering. The three-dimensional (3D) images of depth profiles of Li, $Li_2O$, Nb, and NbO in Figure 1c exhibit more visual images of the vertical distribution of various species, showing the distribution of various species in 3D space. The results above indicate that during Ar-ion sputtering, Li undergoes significant migration and accumulates on the crystal surface.

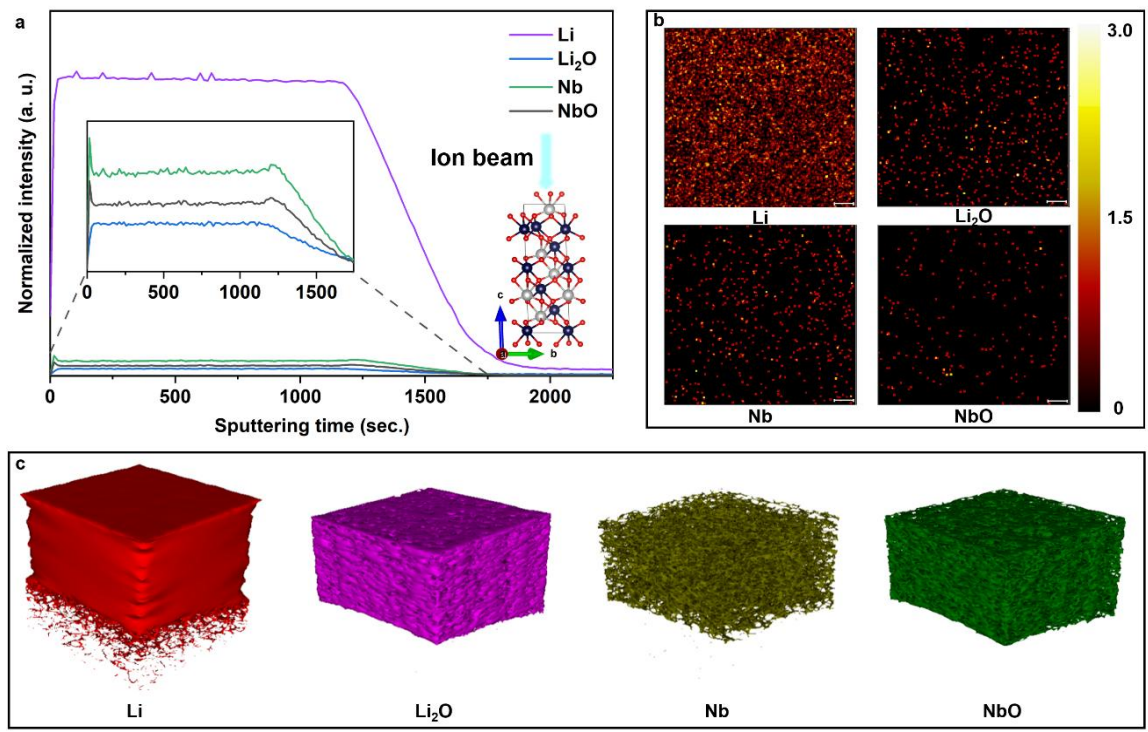

Fig. 1. TOF-SIMS analysis of LN after being sputtered 2250 seconds in positive and negative modes. (b) Overlapped ion mapping images of chemical species. (a) TOF-SIMS depth profiles of various chemical species. (c) Three-dimensional images of depth profiles of various chemical species.

We further studied the defect structure in LN single crystal thin film using spherical ACTEM. Samples were prepared using focused ion beam (FIB)

(Supplementary Fig. S2) and analyzed using ACTEM to investigate the structural defects in LN single crystal thin film. Two FIB samples were prepared in different directions to obtain cross-sectional samples along the [241] zone axis. Regions of different thicknesses were selected for probing, as shown in Fig. 2a-b. We observed some Li columns with increased intensity, while adjacent Nb columns were weakened. When probing thinner regions, the strengthening of some Li columns became more pronounced, indicating the existence of $Nb_{Li}$-$Li_{Nb}$ antisite defects in LN single crystal thin film. We further investigated the concentration of $Nb_{Li}$-$Li_{Nb}$ antisite defects in LN single crystal thin film using position-averaged convergent beam electron diffraction (PACBED) for thickness measurement. The red circles in Fig. 2b represent atomic columns detected by PACBED (Fig. 2c). By combining experimental and simulated data, we determined the thickness of the region to be ~3 nm, i.e., each Nb and Li atomic column contains 9 atoms. By probing four regions (Supplementary Fig. S3), we obtained an average concentration of 0.43(8)% for antisite defects in LN single crystal thin film. We analyzed the atomic surface density of LN single crystal thin film using ACTEM, as shown in Fig. 2d-f. Based on the atomic surface density of LN single crystal thin film, we observed that the distribution of Li atoms in LN crystals is relatively regular, but the surface density maps of Nb and O atoms indicate a decrease in their distribution regularity within the crystal. These results suggest that Nb antisite defects cause irregular distributions within the crystal structure, while the occupation of Li sites by Nb results in distortion of the surrounding O atoms, leading to irregularities in the atomic surface density of O.

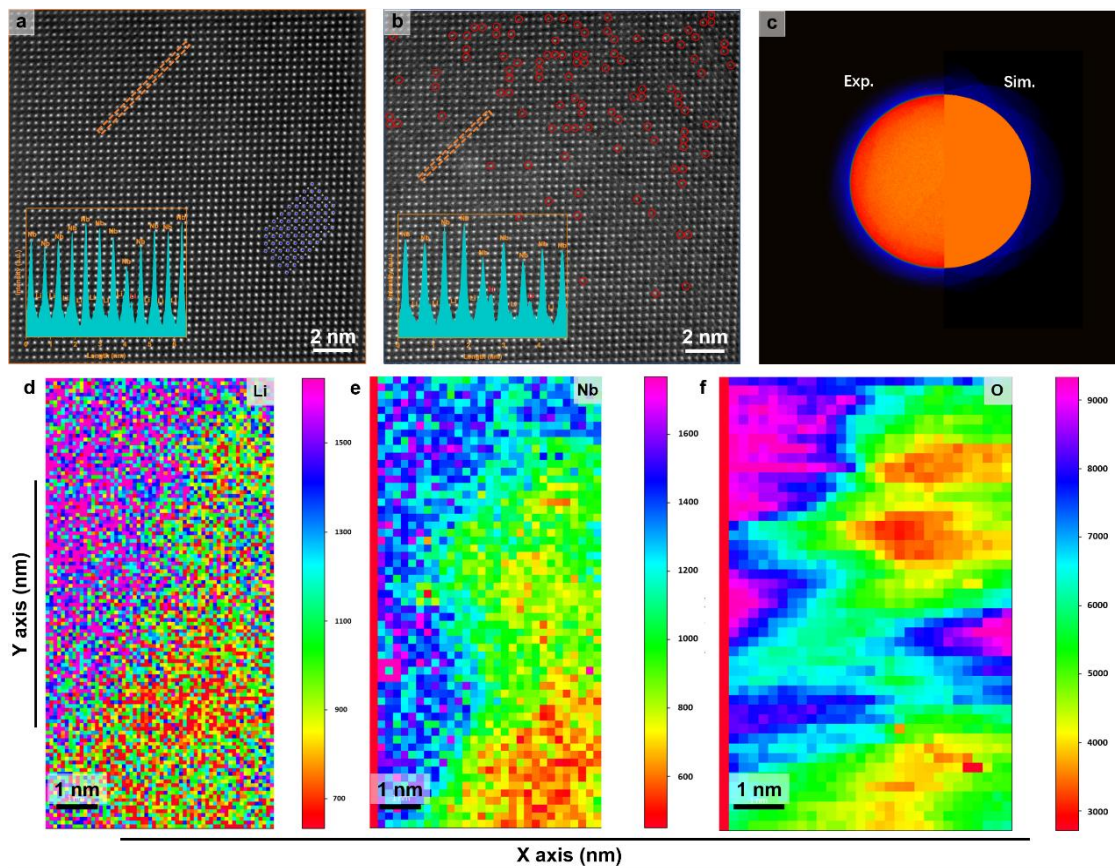

Fig. 2. Characterization of defects in LN single crystal thin film by using aberration-corrected electron microscopy: (a) high-angle annular dark-field (HAADF) image of the thick region, the illustration in the lower left corner represents intensity profile of Li and Nb columns (the orange dashed box indicates the intensity scanning region). (b) HAADF image of the thin region, the illustration in the lower left corner represents intensity profile of Li and Nb columns (the orange dashed box indicates the intensity scanning region). Red circle indicates the PACBED detection area (Estimated concentration for the $Nb_{Li}$-$Li_{Nb}$ antisite pairs: 0.49%). (c) PACBED for thickness measurement, thickness for this region is ~3nm, i.e. each Nb and Li atomic column contains 9 atoms. (d-f) Areal density mapping of Li, Nb and O atoms in LN single crystal thin film.

To further investigate the microstructure of LN single crystal thin film and the coordination environment of Nb, X-ray absorption near-edge structure (XANES) and extended X-ray absorption fine structure (EXAFS) analyses were performed on LN single crystal thin film and reference materials. According to the Nb K-edge XANES spectrum of LN single crystal thin film (Fig. 3a), the absorption threshold of LN single crystal thin film is located after the Nb foil and $Nb_2O_5$. The magnified pre-edge features in Figure 3a indicate a redshift in the absorption edge of Nb in LN single crystal thin film, suggesting an increased oxidation state of Nb in defective LN single crystal thin film. This can be attributed to the substitution of Nb in Li sites, resulting in an increased Nb content in the system, which is consistent with the results from ACTEM. Fig. 3b and 3c depict the near-edge XANES spectra of Li in LN single crystal thin film and the reference $Li_2CO_3$, as well as the absorption spectrum of O. The similarity between the near-edge XANES spectra of Li in LN single crystal thin film and $Li_2CO_3$ suggests that they share a similar coordination environment. The coordination environment was further analyzed using Fourier-transformed extended X-ray absorption fine structure (FT-EXAFS) with $k^3$ weighting at the Nb K-edge (Fig. 3 and Supplementary Fig. 4). The main peak at approximately 1.58 Å (Fig. 3e and 3f) resembles the peak observed in $Nb_2O_5$, which is typically assigned to Nb-O coordination in the first shell. Furthermore, two distinct coordination modes of Nb-O with different bond lengths are observed in LN single crystal thin film, consistent with the structural model. The FT-EXAFS spectrum of LN single crystal thin film exhibits significant long-range reflections, such as Nb-Nb bonding. EXAFS fitting reveals two different coordination numbers for Nb-O in the first shell of Nb, estimated as 2.5±0.5 and 3.6±1.1 (Fig. 3c; detailed information can be found in Supplementary Table 1). This suggests that each Nb atom is coordinated with six O atoms at distances of approximately 1.87±0.01 Å and 2.14±0.01 Å. However, due to the presence of defects, the number of O atoms forming long and short Nb-O bonds may vary, indicating changes in the coordination near the defect sites caused by Nb occupying Li sites. Wavelet transform (WT) analysis was further conducted to study the Nb K-edge EXAFS oscillations of LN single crystal thin film and reference materials. As shown in Fig. 3g-i, the WT analysis of the Nb K-edge reveals a maximum intensity at approximately 5.4 Å$^{-1}$ for LN single crystal thin film, which is close to the weak peak of $Nb_2O_5$ (4.9 Å$^{-1}$), but different from the strongest peaks of Nb foil (7.7 Å$^{-1}$) and $Nb_2O_5$ (9.1 Å$^{-1}$). Additionally, LN single crystal thin film exhibits a secondary peak at approximately 9.8 Å$^{-1}$, similar to the strongest peak of $Nb_2O_5$ (9.1 Å$^{-1}$). These results, in conjunction with the findings from TOF-SIMS and ACTEM, collectively confirm the presence of $Nb_{Li}$ antisite defects in LN single crystal thin film.

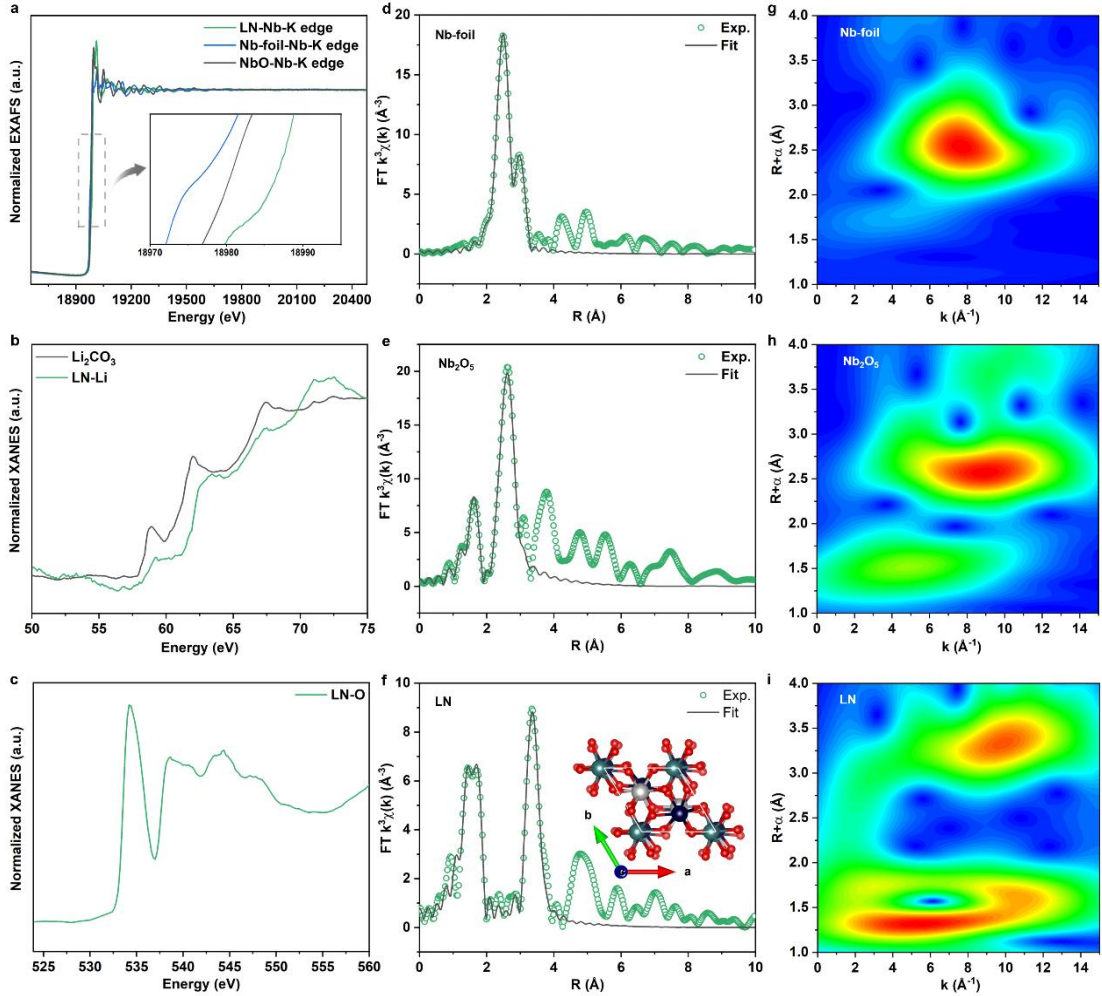

Fig. 3 Local structural characterizations of LN single crystal thin film. (a) Nb K-edge EXAFS spectra, (b) Li and (c) O XANES spectra of LN single crystal thin film and reference samples. (d), (e), and (f) The fitting of the FT R-space Nb K-edge EXAFS of LN single crystal thin film and reference samples, illustration in (f) is a schematic diagram of the structure of LN single crystal thin film, the gray represents Li, the dark purple represents Nb, the red represents O, and the green represents $Nb_{Li}$. (g–i) Wavelet transform for the $k^3$-weighted EXAFS of LN single crystal thin film and reference samples (Nb foil, $Nb_2O_5$, and LN).

**Electron-phonon interactions.** Structures with defects are often more prone to symmetry breaking, as defects disrupt the perfect symmetry of the structure[37,38]. Symmetry refers to the property of a physical system remaining unchanged under certain transformations, such as mirror reflection, rotational symmetry, and translational symmetry. If a structure possesses a certain type of symmetry, its properties remain unchanged under such symmetry transformations. The presence of defects in a structure typically affects its symmetry. For example, if a crystal contains defects, its crystal structure will undergo distortion, thereby breaking the symmetry of the crystal. Similarly, if an object's shape is not completely symmetrical, its symmetry will also be affected. Symmetry breaking can lead to important physical phenomena such as phase transitions and spin glass transitions. Therefore, studying symmetry breaking is crucial for understanding the properties and behavior of materials.

Angle-resolved Raman spectroscopy is a powerful technique for investigating symmetry breaking and related quantum effects in materials[39,40]. In this work, LN presents itself as a promising candidate for exploring these phenomena due to its unique structural, electronic, and optical properties. LN is a ferroelectric material with a perovskite-derived structure, which exhibits a range of intriguing properties, such as large piezoelectric coefficients, strong electro-optic effects, and substantial nonlinear optical coefficients. These properties make LN an attractive material for various applications, including optoelectronic devices, sensors, and quantum information processing. Symmetry breaking in LN is associated with its ferroelectric phase transition, which results from displacements of Li and Nb ions in the crystal lattice, leading to a change in the local symmetry. This change is predicted to be accompanied by the emergence of novel quantum effects, such as entangled photon pairs or topological states, which could be exploited for quantum technologies. Angle-resolved Raman spectroscopy can provide valuable insights into the symmetry breaking and quantum effects in LN. By measuring the Raman scattering as a function of the incident and scattered light angles, this technique can reveal information about the phonon dispersion and the local symmetry changes in the material. Furthermore, angle-resolved Raman spectroscopy can directly probe the coupling between the lattice vibrations and the electronic excitations (i.e. electron-phonon coupling), which is crucial for understanding the underlying mechanisms of the observed quantum effects.

We performed angle-resolved Raman spectroscopy on LN single crystal thin film (as shown in Fig. 4). The tests were carried out at excitation wavelengths of 532 nm and 633 nm, in VH and VV modes. The Raman spectroscopy indicates the presence of electron-phonon coupling in the material. When there is interaction between electronic and lattice vibrations, they exchange energy. The strength of this interaction depends on the excitation energy and the material properties. We compared the symmetry and intensity of Raman-active modes excited at 633 nm (red light) and 532 nm (green light) in LN single crystal thin film, and observed that most Raman modes have enhanced intensities at 633 nm excitation. Under both excitation wavelengths, most of the vibrational modes were the same, but the $A_2T(O_3)$ Raman vibrational mode was excited at 420 cm$^{-1}$ by 532 nm light, and the $ET(O_7)$ Raman vibrational mode was excited at 432 cm$^{-1}$ by 633 nm light. The $A_2T(O_3)$ Raman vibrational mode is infrared-active, and is a one-dimensional (1D) vibration, while the $ET(O_7)$ Raman vibrational mode is both infrared-active and Raman-active, and is a two-dimensional (2D) vibration. This indicates that with the increase of laser intensity, the symmetry center of the LN crystal is broken, and the vibration mode changes from 1D to 2D. lWhen light scatters into the LN single crystal thin film, photon-phonon interactions generate Raman scattering. During angle-resolved polarized Raman tests, the sample was irradiated to produce Raman scattering, and the relative direction between incident light and scattered light was changed by rotating the sample and polarizer. Through the analysis of the test results, we found that the Raman spectrum of LN single crystal thin film has a strong angular dependence, indicating that polarized light breaks the symmetry of the LN single crystal thin film. Through angle-resolved Raman spectroscopy, we can identify changes in electron-phonon interactions.

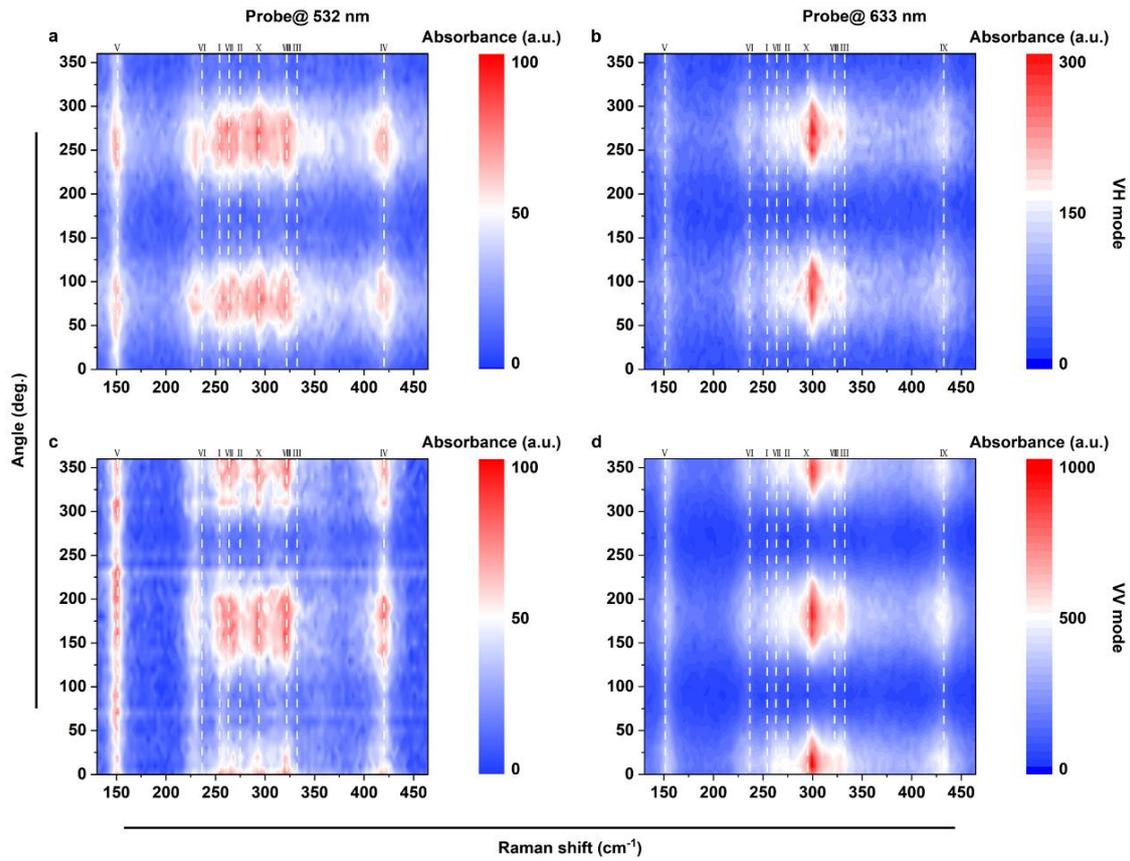

Fig. 4. Angle-resolved Raman spectroscopy measurement probe at 532 nm and 633 nm, respectively. The configuration for angle-resolved Raman spectroscopy involves fixing the laser polarization along the y-axis (designated as $V_L$), setting the direction of the analyzer as $V_R$ or $H_R$, and rotating the sample clockwise around the z-axis, labeled as VH and VV mode, respectively. From the results of angular resolved Raman spectroscopy, it can be seen that the Raman characteristic peaks of lithium niobate have strong angular dependence, indicating that symmetry is broken in the crystal of lithium niobate during the interaction with the polarized light. In addition, between 400 cm$^{-1}$ and 450 cm$^{-1}$, the vibrational modes of LN excited by 532nm and 633nm lasers are different. The Raman numerals at the top of the Raman spectrum represent different Raman vibrational modes, which correspond as follows: I: $A_1T(O_1)$, II: $A_1T(O_2)$, III: $A_1T(O_3)$, IV: $A_2T(O_3)$, V: $ET(O_1)$, VI: $ET(O_3)$, VII: $ET(O_4)$, VIII: $ET(O_5)$, IX: $ET(O_7)$, X: $EL(O_4)$.

We further verified the electron-phonon interaction through transient absorption (TA) measurements. Fig. 5a, d shows the pseudo color map of differential absorption (ΔA) as a function of delay time with 300 nm excitation (pump fluence of 2.7 µJ/cm$^2$). By analyzing the TA test results, LN single crystal thin film shows positive excitonic bleaching at 358 (Fig. 5b) and 722 nm(Fig. 5e). In addition, there is an extra negative peak located at 385 nm (Fig. 5c) and an extra positive peak located at 751 nm (Fig. 5f). To further verify whether ultrafast lasers cause electron-phonon interactions in LN single crystal thin film, we further analyzed the TA data and extracted the oscillation components of TA at 358 nm, 385 nm, 491 nm, 722 nm and 751 nm (Supplementary Fig. 5). By analyzing the oscillation component, we found that the oscillation component at all four wavelengths has strong oscillations with large amplitudes, and the phase changes over time. To prove that there is electron-phonon interaction at all four excitation peaks, we further performed the Fourier transform (Fig. 5g). Through the Fourier transform results, we found that the phonon oscillations at 722 nm and 751 nm are more obvious, especially at 722 nm. At the same time, according to the TA kinetics spectrum, the rising on-site at 385 nm, 722 nm, and 751 nm is delayed (Fig. 5h). All these results prove that under the action of ultrafast lasers, electron-phonon coupling occurs in LN single crystal thin film, that is, electrons and lattice degrees of freedom are coupled. Fig. 5i illustrates the photo-induced exciton dynamics model of LN single crystal thin film. The rise time of the signal at 358 nm is related to the generation of free excitons (FEs) at ~253 fs[41]. The intrinsic recombination of free excitons occurs below 320 ps, featuring a slow decay component $\tau_3$ of FEs[42]. The ultrafast ($\tau_1$) and fast decay ($\tau_2$) time constants at the probing wavelength of 751 nm might be associated with a fast trapping transfer from FEs to self-trapping states (STEs)[42]. The formation of STEs is induced by the relaxation of free excitons (FEs) coupled with longitudinal optical phonons (~0.6 ps) and acoustic phonons (~12 ps)[43]. The ultra-long-lifetime component ($\tau_3$, >1 ns) is assigned to the radiative recombination of STEs[41,42,44]. In general, FEs are formed within ~253 fs after photoexcitation and part of FEs decay through intrinsic radiative recombination for photoluminescent (PL) below 320 ps. Some other FEs are trapped forming STEs in a a time range of 0.6~12 ps, followed by the radiative decay of STEs with an extended PL lifetime > 1 ns.

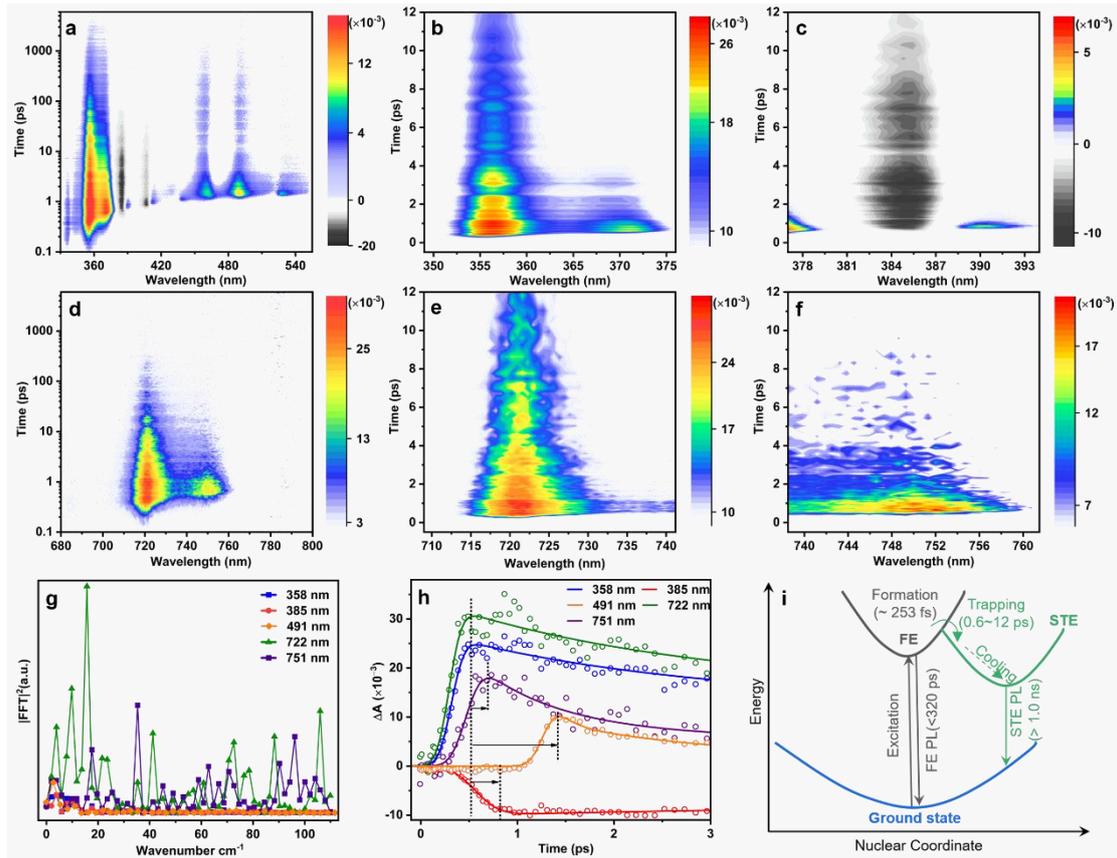

Fig. 5. (a-f) Represent 2D pseudo color map of the fs-TAS profile with 400 nm excitation. (g) FFT power spectrum of the oscillatory components probed at 358 nm, 385 nm, 491 nm, 722 nm, and 751 nm. (h) Difference of onset time. (i) schematic illustrations of photo-induced exciton dynamics model of LN single crystal thin film.

STEs can be generated by a strong coupling between electron and phonon (EPC) transiently and elastically deforming crystal lattice in soft semiconductors upon photoexcitation[45]. The origin of STEs can be from direct optical excitation or the dynamic relaxation of band-edge exciton or carriers[46]. The ultrafast relaxation of energetic carriers through EPC for polar crystals proceeds mainly through a mechanism of LO phonon cascade emission ( Fröhlich interactions) at room temperature[47,48], featured by short-range interactions[49]. A transition from a free exciton to STS occurs with a higher EPC strength[49].

STS is featured by a broad subgap absorption and a formation time in a range of hundreds of femtosecond picoseconds (1D material) to nanoseconds (3D material)[44,45,50-53], depending on the dimensionality of materials and corresponding trapping potential barriers[53]. Besides, the formation of STSs is accompanied by coherent phonon oscillation, which can stabilize STSs through lattice deformation[47,53]. Unlike carriers trapped at permanent defects, STSs are transient defects stabilized by elastic lattice deformation due to EPC[53]. The polar Fröhlich interactions through longitudinal optical (LO) phonon emissions dominate at moderate carrier densities in the visible region, with a LO phonon lifetime of 0.6 ps[52]. The formation time of STEs is comparable to the vibrational period of relevant phonon modes for the 1D system at a time scale of hundreds of femtoseconds[53]. Moreover, radiative recombination of STEs shows an extended lifetime to nanoseconds, in contrast to < 500 ps for free excitons[41].

TA results enable the identification of STEs and STE-related phonon dynamics. Even if the transient absorption spectrum of the studied sample exhibits a complicated spectral feature, the phonon-assisted STSs, through the transition of free excitons, can still be identified in the region between 730 nm to 760 nm. In general, positive bands ($\Delta A > 0$) of TA spectra in the visible region can be attributed to free exciton/carrier absorption, photoinduced bandgap renormalization, or photoinduced absorption (PIA) by the collision between free electrons with LO phonons (Fröhlich interaction, also called as inverse bremsstrahlung absorption) to relax free excitons[47,52]. Besides, the negative bands ($\Delta A < 0$) can arise from Pauli blocking, stimulated emission, or bandgap renormalization[54]. Photobleaching induced by Pauli blocking resonates with the excitation frequency, so only one single or doublet negative peak is observed[54]. Different from Pauli blocking, the photoinduced bandgap renormalization with the presence of photoexcited carriers in the low carrier-density regime is featured by a series of simultaneous photobleaching and corresponding red-shifted PIA sidebands, which intensity changes within 100 ps[54].

As shown in the TA plots of the UV-photoexcited sample (Supplementary Fig. 6), a series of positive and negative bands are observed across the visible region. Note that the detection window between 600 to 700 nm is removed due to the signal contamination from pump pulse scattering (Supplementary Fig. 6c).

As presented in Supplementary Fig. 6, after UV-photoexcitation, a series of negative and positive bands appear alternatively and almost simultaneously throughout the regions between 370 to 550 nm within ~50 ps, which might be originated from photoinduced bandgap renormalization in addition to Pauli blocking (photoinduced bleaching) and band-filling effect[46,54]. In addition, the narrow positive band at 358 nm might be originated from free carrier absorption.

Moreover, the broad positive bands occupying the region between ~700 to 760 nm are highly possible due to the STSs, which develop within 0.6 ps and be peaked at 722 and 751 nm. The almost synchronous variations for doublet peaks at 358/368 and 722/751 cm$^{-1}$ within a time delay of 5 ps demonstrate that the STE stems from the relaxation of free excitons. Additionally, the slight blue shift of hot STSs in Figure 1c within 5 ps indicates the cooling process of hot STEs[41].

With requiring accumulation of free carriers to collide with lattices, the appearance of STSs is expected to exhibit some delay after excitation[47]. The formation times of EPC from free carrier relaxation occur on the picosecond or sub-picosecond time scales relying on Frohlich interaction strength and LO phonon energies[55,56]. Typical formation time (obtained through time constants from decay curve fitting) of ~0.3 ps to ~1.4 ps have been observed for some 2D materials[55,56].

As presented in Fig. 5h and Supplementary Table 2, the onset time (difference in the rising edge between STS and exciton at 358 nm) is ~139 fs at 722 nm, similar timescale to the reported ones[41,44,51]. The formation time correlates with one vibration cycle of phonon[44]. And the phonon energy can be estimated by dividing reduced Planck's constant over the formation time of STS (~139 fs). The phonon energy is 29.8 meV (~241 cm$^{-1}$).

A tri-exponential fitting model was employed to fit decay kinetics at various probing wavelengths (Supplementary Fig. 7). As reported, the photoexcited electrons can be relaxed through rapid electron-electron thermalization ($\tau_1$, ~0.2-0.3 ps), relaxation of hot carriers through LO phonon cascade emission

(Fröhlich interaction) ($\tau_2$, ~1.5–3.5 ps), and electron relaxation through (non)radiative recombining with holes ($\tau_3$, ~5.0–150 ps)[47]. Another study attributes the ultrafast component (0.25 ps) and fast component (4.7 ps) to the contributions of carrier relaxation via LO phonon with Fröhlich interaction and electron-acoustic phonon coupling, respectively[43].

This study uses ultrafast and fast decay components with time constants of $\tau_1$ and $\tau_2$ At 722 and 751 nm are decreased as low as 0.60 ps and 11.75 ps, respectively, in contrast to 385 nm (9.83 and 31.03 ps). Notably, the time constant $\tau_1$ (0.60 ps) at 751 nm is consistent with its corresponding rise time (0.32 ps). The shortened time constants of ultrafast and fast components are due to the presence of LO phonon or acoustic phonon-assisted relaxation of energetic carriers, which facilitates the decay process and shortens decay time constants.

In contrast, the slow decay components of a time constant $\tau_3$ At a probing wavelength of 751 nm is extended to ~1.02 ns. Such a prolonged life is due to the reduced vibrational wavefunction overlap of ground and excited states under the influence of the strong EPC effect[41,44].

The displacive excitation of coherent phonon and impulsive stimulated Raman scattering upon photoexcitation can elucidate transient EPC dynamics[43,44].

Fig. 5g shows a strong oscillation of Fast Fourier transformed ΔA (residual difference) for 722 and 751 nm in contrast to 358 and 385 nm. For 722 and 751 nm, their oscillations are almost overlapped. A series of vibrational modes are identified at frequencies of 3 cm$^{-1}$, 18 cm$^{-1}$, 36 cm$^{-1}$, 63 cm$^{-1}$, 72 cm$^{-1}$, and 97 cm$^{-1}$. The electron relaxation rate varies with different polar phonon modes[47]. Therefore, we can correlate phonon oscillations with decay kinetics. The lower frequency of ~3 cm$^{-1}$ can be translated into a time unit of 11 ps, consistent with $\tau_2$ (~11 to 14 ps) for 722 and 751 nm. Therefore, the frequency of ~3 cm$^{-1}$ is attributed to acoustic phonons modes coupled to electrons in STSs[43,47]. The other frequencies between 18 to 97 cm$^{-1}$ are correlated with LO phonons coupled to electrons in STSs, resulting from certain chemical bond stretchings[53].

**Theoretical analysis**. To investigate the origin of electron-phonon coupling in LN single crystal thin film, we selected a perfect crystal model (LN) and established a model with Nb antisite defects (LN-Nb$_{Li}$) based on the experimentally determined defect concentration from previous studies. DFT calculations were performed on these two models to simulate the electronic structure of LN (Fig. 6). As shown in Figure 6a, we observed significant distortions in the clusters near the defect sites when Li was replaced by Nb. Furthermore, a comparison of the bond lengths in the LN and LN-Nb$_{Li}$ models (Supplementary Table 3) revealed pronounced changes in the crystal structure upon the formation of Nb antisite defects, indicating their impact on the surrounding coordination environment. Our calculations showed evident splitting of the d orbitals of Nb, indicating a modification in the hybridization of clusters caused by Nb occupying the Li site. Additionally, we calculated the band structures of the LN and LN-Nb$_{Li}$ models. Figure 6b displays the projected band structure based on the LN model, and analysis indicated that the conduction band is primarily occupied by Nb, while the valence band is mainly occupied by O. Further analysis of the orbital-projected band structure (Supplementary Fig. 8) revealed that the conduction band originates mainly from the Nb-d orbitals, with a small contribution from the O-*p* orbitals, suggesting hybridization between the O-*p* and Nb-*d* orbitals. The valence band is primarily contributed by the O-*p* orbitals, with a minor contribution from the Nb-*p* orbitals. Figure 6c shows the projected band structure based on the LN-Nb$_{Li}$ model, and analysis revealed that Nb predominantly contributes to both the conduction and valence bands, with a minor contribution from O. Further analysis of the orbital-projected band structure (Supplementary Fig. 9) indicated that the valence band is mainly influenced by the hybridization between the O-*p* and Nb-*d* orbitals, while the conduction band is primarily contributed by the Nb-*d* orbitals, with only small contributions from the O-p orbitals.

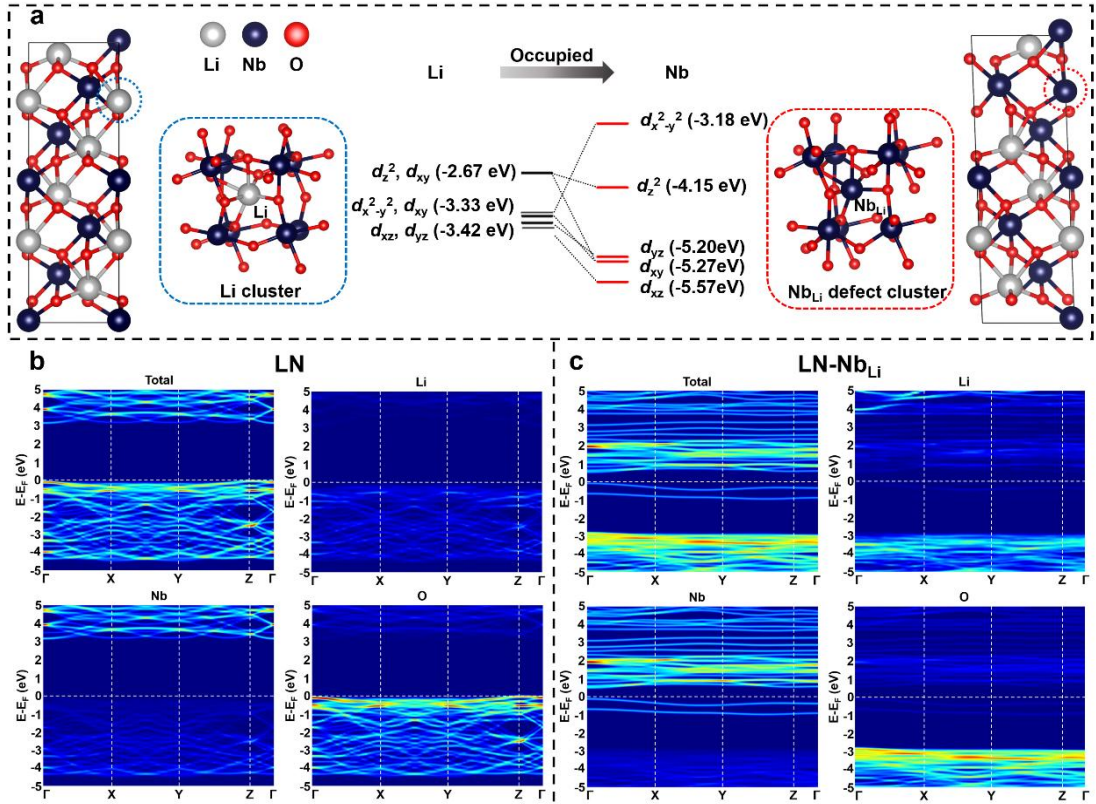

Fig. 6. Simulated electronic structure of the LN crystal, (a) structural analysis and Nb d-orbital splitting analysis based on the LN and LN-Nb$_{Li}$ models, (b) projected band structure based on the LN model, showing the overall projected band structure as well as the projected bands of Li, Nb, and O, and (c) projected band structure based on the LN-Nb$_{Li}$ model, showing the overall projected band structure as well as the projected bands of Li, Nb, and O. Based on the projected band structures, the contributions of each element to the electronic band structure are analyzed.

To elucidate the source of electron-phonon coupling in LN single crystal thin films, we analyzed the crystal structure and the valence band maximum (VBM) and conduction band minimum (CBM). As shown in Fig. 7a and 7b, we observed that the substitution of Nb for Li in the LN crystal resulted in a displacement of Nb$_{Li}$ relative to the lattice along the -a direction (Fig. 7b). Furthermore, we found that upon the formation of the antisite defect, the Li and Nb atoms in the middle of the unit cell along the c direction also experienced displacements. This indicates that the replacement of Li by Nb and the bridging oxygen's influence led to distortions of the Li and Nb atoms in the unit cell, representing the evolution of the defect clusters. Further analysis of the 2D slices of the CBM and VBM in the defect atomic layer revealed the following: in the LN model (Fig. 7c), the CBM was primarily contributed by Nb, while the VBM was mainly influenced by Nb and O, with no significant contribution from the Li at the antisite defect position. In the LN-$Nb_{Li}$ model (Fig. 7d), after Nb replaced Li to form the $Nb_{Li}$ antisite defect, O exhibited some contribution to the CBM, and both Nb and O showed more pronounced contributions to the VBM. This indicates that the formation of the $Nb_{Li}$ antisite defect led to distortions in the O clusters near the defect site and further hybridization between the orbitals of Nb and O. Moreover, the formation of the antisite defect induced lattice distortions, resulting in electron-phonon coupling in LN single crystal thin films.

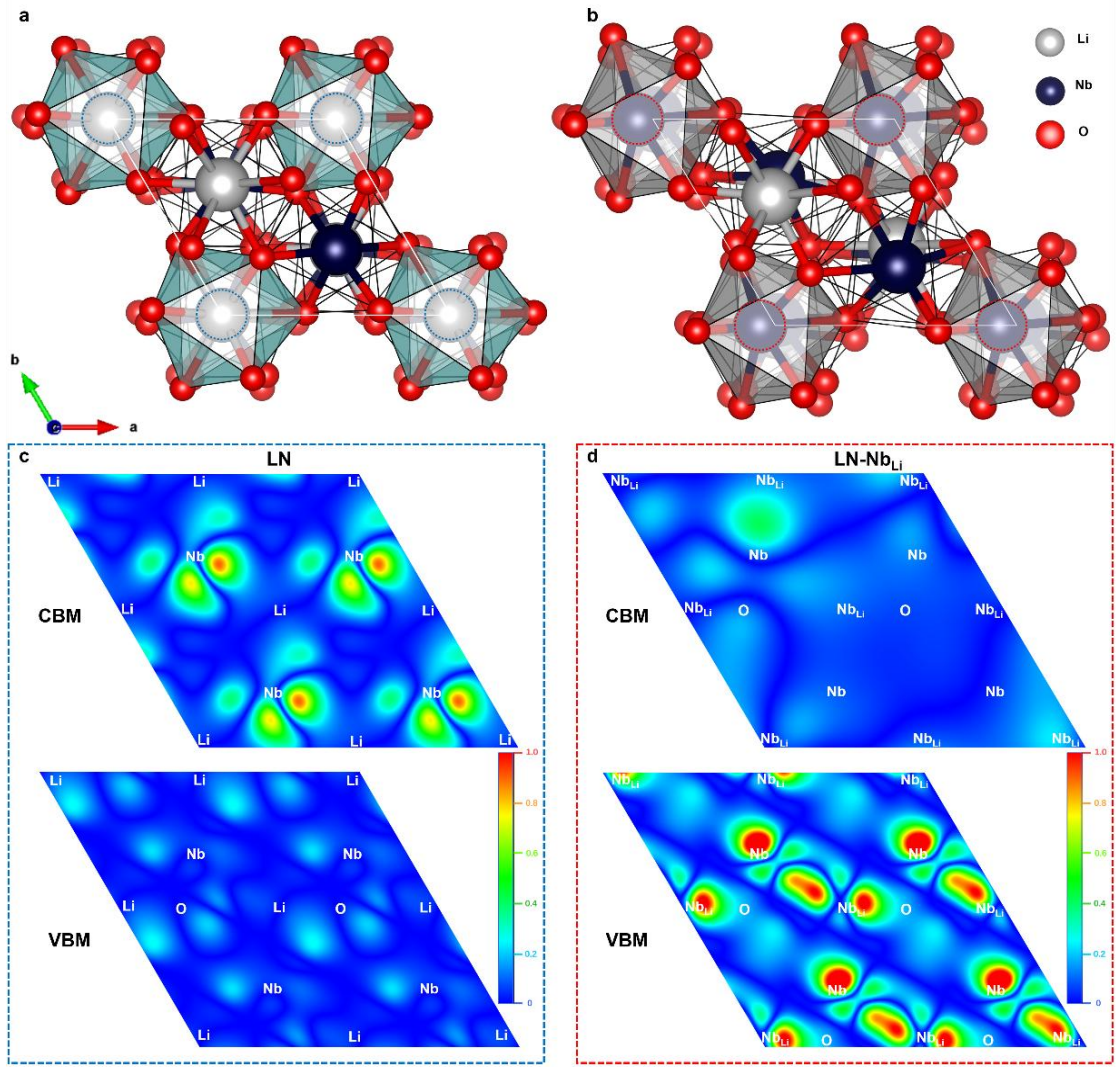

Fig. 7. (a) Illustration of the Li-O polyhedron in the LN model, where blue circles indicate the Li sites to be replaced by Nb. (b) Illustration of the Nb antisite defect N-O polyhedron in the LN-Nb$_{Li}$ model, with the NbLi antisite defect depicted by red circles. (c) 2D slices of the CBM and VBM in the LN model corresponding to the antisite defect positions of Li. (d) 2D slices of the CBM and VBM in the LN-Nb$_{Li}$ model corresponding to the NbLi positions.

**Conclusion**

We conducted a comprehensive, multi-level, and multi-scale analysis of the defect types and defect concentrations in LN single crystal thin films by combining various structural analysis techniques, including TOF-SIMS, ACTEM, and XAS. Furthermore, we observed the phenomenon of electroacoustic coupling in LN single crystal thin films through angle-resolved Raman spectroscopy and TA measurements and investigated the mechanism of electroacoustic coupling through DFT calculations. The significance of LN in the field of photonics is gradually emerging, with a trend towards replacing Si, indicating its potential as a quantum material and for quantum device applications. LN single crystal thin films on insulating substrates serve as a new integrated photonics platform, and our work demonstrates the presence of electron-phonon coupling, which is crucial for its application in the field of quantum materials and coupled controllable quantum devices. Therefore, this study provides a research foundation for exploring quantum technologies based on LN crystals and holds significant practical implications for advancing the application of LN crystals as quantum materials.

**Methods**

**LN single crystal thin films.** A z-cut LN single crystal thin films wafer (thickness of 300 nm) was prepared by Pro. Hui Hu *via* high-dose-ion-implantation, which was used in the characterization (described in Supplementary Materials, 1. Lithium niobate (LN) large-sized single crystals and LN single crystal thin films.).

**Experimental details.** Depth analysis of the LN single crystal thin films structure was performed using TOF-SIMS with an ULVAC-PHI instrument from Japan. The experimental conditions were as follows: Initial ion beam: $Bi_3^{++}$LMIG, initial ion beam energy: 30 keV, test area: 50 μm × 50 μm, sputtering ion beam: argon gas gun, sputtering ion beam energy/current: 4 kV/200 nA, sputtering area: 200 μm × 200 μm, relative silicon dioxide ($SiO_2$) sputtering rate: 1.0 nm/s, relative sample sputtering rate: 0.25 nm/s. HAADF images were obtained using a Thermalfisher scientific titan themsis Z instrument. The samples were prepared by Focused Ion Beam (FIB) at 2 kV using an FEI Scios2 instrument. XAS including XANES and EXAFS of the Nb K-edge was collected at the SPring-8 14b2 facility, where a pair of channel-cut Si(111) crystals served as the monochromator. The storage ring operated at an energy of 8.0 GeV with an average electron current of 99.5 mAThe XANES of Li and O K-edge were collected at the Singapore Synchrotron Light Source (SSLS) center, using a pair of channel-cut Si(111) crystals in the monochromator. The storage ring operated at an energy of 2.5 GeV with an average electron current below 200 mA. The Raman spectroscopy used for angular resolved Raman was performed using a Horiba Jobin Yvon LabRam HR800 confocal micro-Raman spectrometer. The system was equipped with a liquid nitrogen-cooled CCD detector and an Olympus 100x objective (numerical aperture of 0.90). The grating had a line density of 1200 lines/mm. The laser sources used were a 532 nm laser from a Nd:YAG laser and a 633 nm laser from a He-Ne laser. TA measurements were conducted using the Ultrafast Helios system (320-1600 nm, 1 kHz high-sensitivity detection system). The specific parameters were as follows: The laser source was a Coherent Astrella laser with a pulse energy greater than 7 mJ, a central wavelength of 800 nm, a pulse width less than 100 fs, and a repetition rate of 1 kHz. The optical parametric amplifier used was an OPerA-Solo, with a wavelength tuning range of 240-2600 nm.

**Calculation methods.** The present calculations utilized the Vienna ab initio Simulation Package (VASP)[57,58], which implements Density Functional Theory (DFT) along with the projector-augmented-wave (PAW) formalism. In this approach, the Li $2s^1$, Nb $4p^6 5s^1 4d^4$, and O $2s^2 2p^4$ states were considered as valence electrons. The electronic wave functions were expanded using plane waves with an energy cutoff of 500 eV. The (semi)local generalized gradient approximation (GGA) of Perdew, Burke, and Ernzerhof (PBE) functional was employed to describe the electron exchange and correlation (XC) effects, optimize the configurations, and calculate the defect formation energies[59]. The force convergence criterion for structural relaxation was set to 0.01 eV/Å. To sample the Brillouin zones, a 4×4×4 Γ-centered Monkhorst-Pack k-point mesh was utilized.

**Acknowledgments**

This work was supported by the National Natural Science Foundation of China

(52203366, 52220105010). K.C. also acknowledges Qilu Young Scholars Program of Shandong University.

## References

1    Li, M. *et al.* Lithium niobate photonic-crystal electro-optic modulator. *Nat Commun* **11**, 4123, doi:10.1038/s41467-020-17950-7 (2020).
2    Buse, K., Adibi, A. & Psaltis, D. Non-volatile holographic storage in doubly doped lithium niobate crystals. *Nature* **393**, 665-668, doi:10.1038/31429 (1998).
3    Wang, C. *et al.* Integrated lithium niobate electro-optic modulators operating at CMOS-compatible voltages. *Nature* **562**, 101-104, doi:10.1038/s41586-018-0551-y (2018).
4    Yu, M. *et al.* Integrated femtosecond pulse generator on thin-film lithium niobate. *Nature* **612**, 252-258, doi:10.1038/s41586-022-05345-1 (2022).
5    Zhang, M. *et al.* Broadband electro-optic frequency comb generation in a lithium niobate microring resonator. *Nature* **568**, 373-377, doi:10.1038/s41586-019-1008-7 (2019).
6    Pohl, D. *et al.* An integrated broadband spectrometer on thin-film lithium niobate. *Nat. Photonics* **14**, 24-29, doi:10.1038/s41566-019-0529-9 (2019).
7    Wei, D. *et al.* Experimental demonstration of a three-dimensional lithium niobate nonlinear photonic crystal. *Nat. Photonics* **12**, 596-600, doi:10.1038/s41566-018-0240-2 (2018).
8    Ge, R., Yan, X., Chen, Y. & Chen, X. Broadband and lossless lithium niobate valley photonic crystal waveguide [Invited]. *Chinese Optics Letters* **19**, 060014, doi:10.3788/col202119.060014 (2021).
9    Wang, C. *et al.* Monolithic lithium niobate photonic circuits for Kerr frequency comb generation and modulation. *Nat Commun* **10**, 978, doi:10.1038/s41467-019-08969-6 (2019).
10   Marpaung, D., Yao, J. & Capmany, J. Integrated microwave photonics. *Nat. Photonics* **13**, 80-90, doi:10.1038/s41566-018-0310-5 (2019).
11   Wang, J., Sciarrino, F., Laing, A. & Thompson, M. G. Integrated photonic quantum technologies. *Nat. Photonics* **14**, 273-284, doi:10.1038/s41566-019-0532-1 (2019).
12   Jin, H. *et al.* Compact engineering of path-entangled sources from a monolithic quadratic nonlinear photonic crystal. *Phys Rev Lett* **111**, 023603, doi:10.1103/PhysRevLett.111.023603 (2013).
13   Boes, A. *et al.* Lithium niobate photonics: Unlocking the electromagnetic spectrum. *Science* **379**, eabj4396, doi:10.1126/science.abj4396 (2023).
14   Jin, H. *et al.* On-chip generation and manipulation of entangled photons based on reconfigurable lithium-niobate waveguide circuits. *Phys Rev Lett* **113**, 103601, doi:10.1103/PhysRevLett.113.103601 (2014).

15      Zhao, J., Ma, C., Rusing, M. & Mookherjea, S. High Quality Entangled Photon Pair Generation in Periodically Poled Thin-Film Lithium Niobate Waveguides. *Phys Rev Lett* **124**, 163603, doi:10.1103/PhysRevLett.124.163603 (2020).
16      Renaud, D. *et al.* Sub-1 Volt and high-bandwidth visible to near-infrared electro-optic modulators. *Nat Commun* **14**, 1496, doi:10.1038/s41467-023-36870-w (2023).
17      Xu, M. *et al.* High-performance coherent optical modulators based on thin-film lithium niobate platform. *Nat Commun* **11**, 3911, doi:10.1038/s41467-020-17806-0 (2020).
18      He, M. *et al.* High-performance hybrid silicon and lithium niobate Mach–Zehnder modulators for 100 Gbit s−1 and beyond. *Nat. Photonics* **13**, 359-364, doi:10.1038/s41566-019-0378-6 (2019).
19      Bartasyte, A., Margueron, S., Baron, T., Oliveri, S. & Boulet, P. Toward High-Quality Epitaxial $LiNbO_3$ and $LiTaO_3$ Thin Films for Acoustic and Optical Applications. *Advanced Materials Interfaces* **4**, 1600998, doi:10.1002/admi.201600998 (2017).
20      Jiang, W. *et al.* Efficient bidirectional piezo-optomechanical transduction between microwave and optical frequency. *Nat Commun* **11**, 1166, doi:10.1038/s41467-020-14863-3 (2020).
21      Yan, Z. W. *et al.* Probing Rotated Weyl Physics on Nonlinear Lithium Niobate-on-Insulator Chips. *Phys Rev Lett* **127**, 013901, doi:10.1103/PhysRevLett.127.013901 (2021).
22      Diddams, S. A., Vahala, K. & Udem, T. Optical frequency combs: Coherently uniting the electromagnetic spectrum. *Science* **369**, eaay3676, doi:10.1126/science.aay3676 (2020).
23      Metcalf, A. J., Torres-Company, V., Leaird, D. E. & Weiner, A. M. High-Power Broadly Tunable Electrooptic Frequency Comb Generator. *IEEE J. Sel. Top. Quantum Electron.* **19**, 231-236, doi:10.1109/jstqe.2013.2268384 (2013).
24      Rueda, A., Sedlmeir, F., Kumari, M., Leuchs, G. & Schwefel, H. G. L. Resonant electro-optic frequency comb. *Nature* **568**, 378-381, doi:10.1038/s41586-019-1110-x (2019).
25      Hu, Y. *et al.* High-efficiency and broadband on-chip electro-optic frequency comb generators. *Nat. Photonics* **16**, 679-685, doi:10.1038/s41566-022-01059-y (2022).
26      Luo, Q., Bo, F., Kong, Y. F., Zhang, G. Q. & Xu, J. J. Advances in lithium niobate thin-film lasers and amplifiers: a review. *Advanced Photonics* **5**, doi:Artn 034002 10.1117/1.Ap.5.3.034002 (2023).
27      B., L. *Now entering, Lithium Niobate Valley*, <https://www.seas.harvard.edu/news/2017/12/now-entering-lithium-niobate-valley> (2017).
28      Zhu, D. *et al.* Spectral control of nonclassical light pulses using an integrated thin-film lithium niobate modulator. *Light Sci Appl* **11**, 327, doi:10.1038/s41377-022-01029-7 (2022).
29      Herter, A. *et al.* Terahertz waveform synthesis in integrated thin-film lithium niobate platform. *Nat Commun* **14**, 11, doi:10.1038/s41467-022-35517-6 (2023).
30      Xu, Y. *et al.* Bidirectional interconversion of microwave and light with thin-film

lithium niobate. *Nature Communications* **12**, 4453, doi:10.1038/s41467-021-24809-y (2021).

31  Yuan, S. *et al.* Strongly Enhanced Second Harmonic Generation in a Thin Film Lithium Niobate Heterostructure Cavity. *Phys. Rev. Lett.* **127**, 153901, doi:10.1103/PhysRevLett.127.153901 (2021).

32  Tokura, Y., Kawasaki, M. & Nagaosa, N. Emergent functions of quantum materials. *Nature Physics* **13**, 1056-1068, doi:10.1038/nphys4274 (2017).

33  Gong, Y. & Gu, L. Degrees of freedom for energy storage material. *Carbon Energy* **4**, 633-644, doi:10.1002/cey2.195 (2022).

34  Kong, Y. *et al.* Recent Progress in Lithium Niobate: Optical Damage, Defect Simulation, and On-Chip Devices. *Adv Mater* **32**, e1806452, doi:10.1002/adma.201806452 (2020).

35  Chen, K. *et al.* Microstructure and defect characteristics of lithium niobate with different Li concentrations. *Inorganic Chemistry Frontiers* **8**, 4006-4013, doi:10.1039/d1qi00562f (2021).

36  Wang, W. *et al.* Interaction between Mo and intrinsic or extrinsic defects of Mo doped LiNbO3 from first-principles calculations. *J Phys Condens Matter* **32**, 255701, doi:10.1088/1361-648X/ab7ada (2020).

37  Suzuki, R. *et al.* Valley-dependent spin polarization in bulk MoS2 with broken inversion symmetry. *Nat Nanotechnol* **9**, 611-617, doi:10.1038/nnano.2014.148 (2014).

38  Cheong, S.-W., Talbayev, D., Kiryukhin, V. & Saxena, A. Broken symmetries, non-reciprocity, and multiferroicity. *npj Quantum Materials* **3**, doi:10.1038/s41535-018-0092-5 (2018).

39  Quan, J. *et al.* Phonon renormalization in reconstructed MoS2 moire superlattices. *Nat Mater* **20**, 1100-1105, doi:10.1038/s41563-021-00960-1 (2021).

40  Wang, J. *et al.* Determination of Crystal Axes in Semimetallic T′-MoTe2 by Polarized Raman Spectroscopy. *Advanced Functional Materials* **27**, 1604799, doi:10.1002/adfm.201604799 (2017).

41  Shi, M. *et al.* Tuning Exciton Recombination Pathways in Inorganic Bismuth-Based Perovskite for Broadband Emission. *Energy Material Advances* **2022**, 9845942, doi:Artn 9845942

10.34133/2022/9845942 (2022).

42  Yang, B. *et al.* Lead‐Free, Air‐Stable All‐Inorganic Cesium Bismuth Halide Perovskite Nanocrystals. *Angewandte Chemie International Edition* **56**, 12471-12475, doi:10.1002/anie.201704739 (2017).

43  Wu, B. *et al.* Strong self-trapping by deformation potential limits photovoltaic performance in bismuth double perovskite. *Science Advances* **7**, doi:10.1126/sciadv.abd3160 (2021).

44  Sui, X. *et al.* Zone-Folded Longitudinal Acoustic Phonons Driving Self-Trapped State Emission in Colloidal CdSe Nanoplatelet Superlattices. *Nano Letters* **21**, 4137-4144, doi:10.1021/acs.nanolett.0c04169 (2021).

45  Tan, J. *et al.* Self-trapped excitons in soft semiconductors. *Nanoscale* **14**, 16394-16414, doi:10.1039/d2nr03935d (2022).

46   Wu, X. *et al.* Trap States in Lead Iodide Perovskites. *Journal of the American Chemical Society* **137**, 2089-2096, doi:10.1021/ja512833n (2015).

47   Glinka, Y. D., Li, J., He, T. & Sun, X. W. Clarifying Ultrafast Carrier Dynamics in Ultrathin Films of the Topological Insulator Bi2Se3 Using Transient Absorption Spectroscopy. *ACS Photonics* **8**, 1191-1205, doi:10.1021/acsphotonics.1c00115 (2021).

48   Fu, J. *et al.* Hot carrier cooling mechanisms in halide perovskites. *Nature Communications* **8**, 1300, doi:10.1038/s41467-017-01360-3 (2017).

49   Yamada, Y. & Kanemitsu, Y. Electron-phonon interactions in halide perovskites. *Npg Asia Materials* **14**, doi:ARTN 48
10.1038/s41427-022-00394-4 (2022).

50   Yang, B. *et al.* Colloidal Synthesis and Charge‐Carrier Dynamics of Cs2AgSb1−yBiyX6 (X: Br, Cl; 0 ≤y ≤1) Double Perovskite Nanocrystals. *Angewandte Chemie International Edition* **58**, 2278-2283, doi:10.1002/anie.201811610 (2019).

51   Luo, J. *et al.* Efficient and stable emission of warm-white light from lead-free halide double perovskites. *Nature* **563**, 541-545, doi:10.1038/s41586-018-0691-0 (2018).

52   Smith, M. D. & Karunadasa, H. I. White-Light Emission from Layered Halide Perovskites. *Accounts of Chemical Research* **51**, 619-627, doi:10.1021/acs.accounts.7b00433 (2018).

53   Hu, T. *et al.* Mechanism for Broadband White-Light Emission from Two-Dimensional (110) Hybrid Perovskites. *The Journal of Physical Chemistry Letters* **7**, 2258-2263, doi:10.1021/acs.jpclett.6b00793 (2016).

54   Pogna, E. A. A. *et al.* Photo-Induced Bandgap Renormalization Governs the Ultrafast Response of Single-Layer MoS2. *ACS Nano* **10**, 1182-1188, doi:10.1021/acsnano.5b06488 (2016).

55   Thilagam, A. Exciton formation assisted by longitudinal optical phonons in monolayer transition metal dichalcogenides. *Journal of Applied Physics* **120**, 124306 doi:10.1063/1.4963123 (2016).

56   Li, C. *et al.* Ultrafast and broadband photodetectors based on a perovskite/organic bulk heterojunction for large-dynamic-range imaging. *Light: Science & Applications* **9**, 31, doi:10.1038/s41377-020-0264-5 (2020).

57   Kresse, G. & Furthmüller, J. Efficient iterative schemes forab initiototal-energy calculations using a plane-wave basis set. *Physical Review B* **54**, 11169-11186, doi:10.1103/PhysRevB.54.11169 (1996).

58   Kresse, G. & Hafner, J. Ab Initio Molecular Dynamics for Open-Shell Transition Metals. *Physical Review B* **48**, 13115-13118, doi:10.1103/physrevb.48.13115 (1993).

59   Perdew, J. P., Burke, K. & Ernzerhof, M. Generalized Gradient Approximation Made Simple. *Phys. Rev. Lett.* **77**, 3865-3868, doi:10.1103/PhysRevLett.77.3865 (1996).

# Supplementary Materials

# Electron-phonon coupling in lattice engineering of lithium niobate single crystal thin films

Guoqiang Shi[1, #], Kunfeng Chen[2, #], Hui Hu[3], Gongbin Tang[2], Dongfeng Xue[1,4,*]

1. Multiscale Crystal Materials Research Center, Shenzhen Institute of Advanced Technology, Chinese Academy of Sciences, Shenzhen 518055, China

2. Institute of Novel Semiconductors, State Key Laboratory of Crystal Materials, Shandong University, Jinan 250100, China

3. School of Physics, State Key Laboratory of Crystal Materials, Shandong University, Jinan 250100, China

4. Shenzhen Key Laboratory of New Information Display and Storage Materials, College of Materials Science and Engineering, Shenzhen University, Shenzhen 518060, China

5. Shenzhen Institute for Advanced Study, University of Electronic Science and Technology of China, Shenzhen 518110, China

\# These authors contributed equally to this work.
Correspondence to: dfxue@uestc.edu.cn

## 1. Lithium niobate (LN) large-sized single crystals and LN single crystal thin films.

The growth of a single crystal of LN was carried out using the Czochralski (Cz) method under ambient air atmosphere. The sintered polycrystalline LN was placed inside a Pt crucible equipped with an intermediate frequency induction heating system and subjected to heating in a Cz furnace until complete melting of the polycrystalline material was achieved. The molten material was held at a constant temperature for a specific duration. The c-direction of LN was employed as the seed crystal, and manual seeding was performed. Subsequent crystal growth processes were controlled by an automated control system. The specific growth parameters employed were a pulling rate ranging from 2.8 to 4 mm·h$^{-1}$ and a rotation rate ranging from 3 to 10 revolutions per minute. Based on the large-sized single crystals of LN grown in our laboratory, we obtained 300nm-thick LN single crystal thin films using the ion implantation technique.

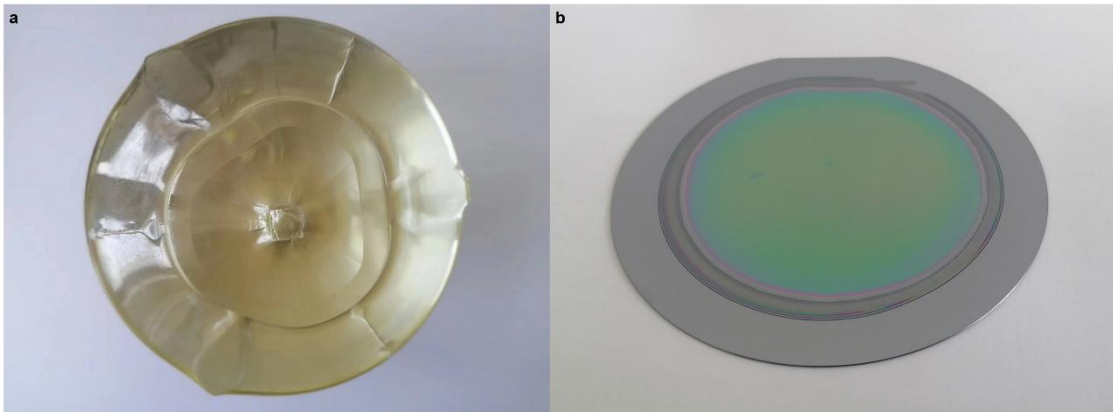

Supplementary Fig. 1| (a) ϕ4-inch LN crystals grown via the fast Czochralski method in our lab, (b) 300nm-thick LN single crystal thin film.

## 2. Fabrication of LN single crystal thin films by FIB.

We utilized focused ion beam (FIB) processing as a technique to precisely fabricate and investigate lithium niobate (LN) single crystal thin films. The primary objective of this approach was to conduct a comprehensive analysis of the defect structures and defect types present within the LN single crystal thin films. By employing FIB, we aimed to determine the specific locations of defects and accurately quantify their concentrations. FIB processing provided us with the ability to precisely modify and manipulate the LN thin films at a microscopic level, allowing us to investigate the intricate details of the defect landscape. This technique allowed us to selectively remove material, create nanostructures, and perform cross-sectional imaging, enabling a thorough examination of the defect characteristics. The characterization and analysis of defects in LN single crystal thin films are of paramount importance in understanding their properties and performance. By elucidating the defect structures and their influence on the material's properties, we aimed to enhance our knowledge of LN thin film growth and optimize their potential applications in various electronic and optoelectronic devices. Through our FIB-based processing and subsequent analysis, we aimed to gain insights into the defect mechanisms, defect densities, and spatial distribution of defects within the LN single crystal thin films. This information serves as a foundation for further investigations and provides valuable guidance for the development of defect engineering strategies to improve the performance and reliability of LN-based devices.

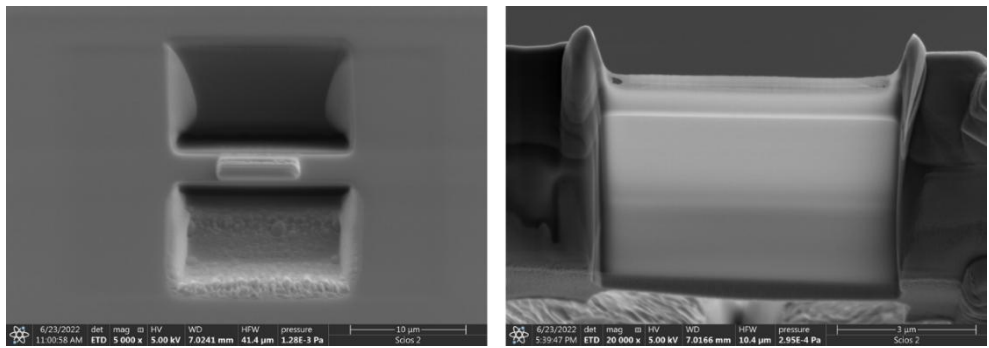

Supplementary Fig. 2| Two FIB specimens were prepared along different directions in order to obtain a cross-section specimen along [241] zone axis.

## 3. Quantification on the concentration of $Nb_{Li}$-$Li_{Nb}$ antisite defects of LN single crystal thin films.

We employed aberration-corrected transmission electron microscopy (ACTEM) to investigate high-angle annular dark-field (HAADF) images of different regions within lithium niobate (LN) single crystal thin films. Additionally, we utilized position-averaged convergent beam electron diffraction (PACBED) in conjunction with HAADF imaging to further analyze the distribution of NbLi-LiNb antisite defects, thereby enabling a more accurate estimation of the defect concentration. The ACTEM technique allowed us to achieve high-resolution imaging of the LN thin films, providing detailed information about the atomic arrangement and defect structures present in different regions. The HAADF imaging mode, in particular, enhanced our ability to discern the atomic column contrasts, facilitating the identification and characterization of defects within the LN thin films. Furthermore, PACBED analysis was performed to gain insights into the distribution and arrangement of NbLi-LiNb antisite defects. By utilizing the convergent electron beam and analyzing the diffraction patterns, we obtained valuable information about the local crystallographic orientation and the presence of antisite defects in the LN thin films. The combination of PACBED with HAADF imaging enabled a comprehensive examination of the defect distribution across the sample. Through the quantitative analysis of the HAADF images and PACBED data, we were able to estimate the defect concentration within the LN single crystal thin films. This information provided valuable insights into the defect formation mechanisms and their impact on the material's properties. Moreover, it served as a foundation for further understanding and optimizing the growth processes of LN thin films, with the ultimate goal of enhancing their performance in various electronic and photonic applications. The ACTEM, HAADF imaging, and PACBED techniques employed in this study offered a powerful approach for characterizing and quantifying defects in LN single crystal thin films. The comprehensive analysis of the defect distribution and concentration provided valuable information for the development of defect engineering strategies, enabling the enhancement of LN thin film properties and the realization of their full potential in advanced device applications.

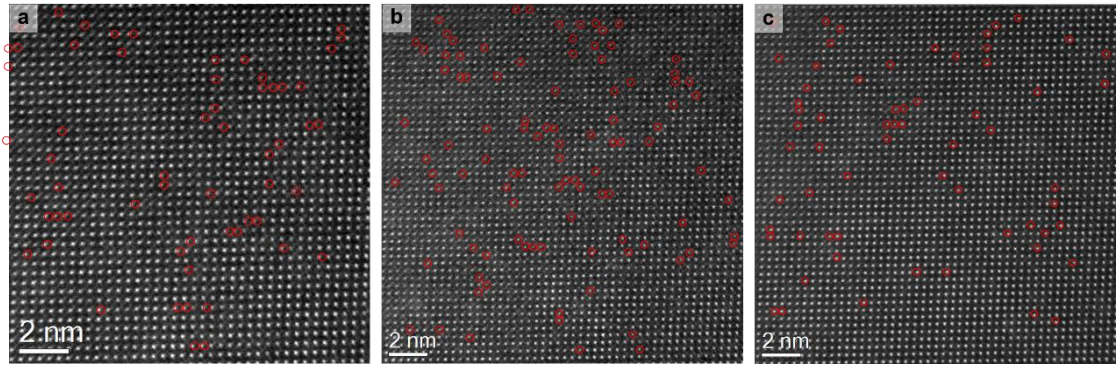

Supplementary Fig. 3| Estimated concentration for the $Nb_{Li}$-$Li_{Nb}$ antisite pairs : (a) 0.46%, (b) 0.47%, and (c) 0.30%.

## 4. Fourier transform magnitude of $k^3$-weighted Nb-K-edge EXAFS spectra.

In the k-space of X-ray absorption spectroscopy (XAS), distinct oscillation patterns can be observed for different coordination elements. Lighter elements exhibit the strongest oscillations at lower wave numbers in the k-space, while heavier elements show the strongest oscillations at higher wave number positions. This phenomenon can be attributed to the differences in the electronic structures and atomic arrangements of the elements. Lighter elements typically have fewer electrons and larger atomic radii, leading to more delocalized electronic states and lower binding energies. Consequently, their oscillatory behavior is more pronounced at lower wave numbers in the k-space. In contrast, heavier elements possess more electrons and smaller atomic radii, resulting in stronger electron-electron interactions and higher binding energies. This leads to oscillations that are more prominent at higher wave number positions in the k-space. The observed variations in the k-space oscillation patterns provide valuable insights into the electronic properties and bonding characteristics of different coordination elements. Understanding these patterns is essential for the interpretation and analysis of XAS data, enabling the identification and characterization of specific elements within a sample and facilitating the investigation of their chemical environments and electronic interactions.

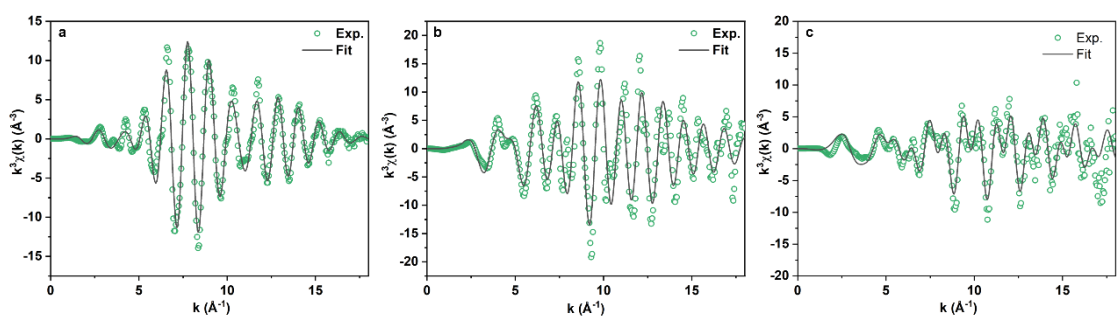

Supplementary Fig. 4| Fitting curves of the $k^3$-weighted Nb K-edge EXAFS for (a) Nb foil, (b) $Nb_2O_5$, and (c) LN crystal.

## 5. Fourier transform magnitude of $k^3$-weighted Nb-K-edge EXAFS spectra.

When performing extended X-ray absorption fine structure (EXAFS) analysis at the Nb K-edge, various fitting parameters are utilized. These parameters are essential for accurately modeling the local atomic structure and investigating the coordination environment of niobium (Nb) atoms. The fitting parameters typically include the coordination number (C.N.), which represents the number of neighboring atoms surrounding the Nb atom. The C.N. provides information about the bonding geometry and the extent of atomic clustering in the sample. In addition to the C.N., the EXAFS fitting also involves the determination of bond lengths (R) between the Nb atom and its neighboring atoms. These bond lengths provide insights into the distances at which the Nb atom interacts with its surrounding atoms, contributing to the understanding of chemical bonding and structural arrangements. Furthermore, the fitting procedure involves the determination of Debye-Waller factors ($\sigma^2$), which account for thermal vibrations of atoms in the crystal lattice. The $\sigma^2$ values indicate the extent of atomic disorder and thermal motion within the local environment of the Nb atom. Other important fitting parameters may include the coordination shell amplitude reduction factors ($S0^2$), which account for multiple scattering effects, and the energy shift ($\Delta E0$), which corrects for any energy calibration errors in the experimental data. By carefully adjusting and optimizing these fitting parameters, it becomes possible to extract detailed information about the local atomic structure, including bond lengths, coordination numbers, and thermal disorder around the Nb atoms. The comprehensive analysis of these parameters plays a crucial role in understanding the physical and chemical properties of Nb-containing materials and aids in the design and optimization of materials with desired functionalities.

Supplementary Table 1. EXAFS fitting parameters at the Nb K-edge($S_0^2$ = 0.99).

| Sample | Path | C.N. | R (Å) | $\sigma^2 \times 10^3$ (Å$^2$) | ΔE (eV) | R factor |
|---|---|---|---|---|---|---|
| Nb foil | Nb-Nb | 8* | 2.86±0.01 | 7.5±0.2 | -7.8±0.4 | 0.001 |
| | Nb-Nb | 6* | 3.31±0.01 | 8.2±0.2 | -4.6±0.8 | |
| NbO | Nb-O | 4.0±0.8 | 2.12±0.01 | 4.3±1.5 | -4.4±2.8 | 0.005 |
| | Nb-Nb | 6.8±0.8 | 2.98±0.01 | 4.5±0.6 | -8.5±1.3 | |
| LN | Nb-O | 2.5±0.5 | 1.87±0.01 | 1.7±0.9 | -3.3±4.1 | 0.011 |
| | Nb-O | 3.6±1.1 | 2.14±0.01 | 7.6±3.1 | | |
| | Nb-Nb | 3.8±0.7 | 3.75±0.01 | 3.8±0.7 | -10.9±1.8 | |

*C.N.*: coordination numbers; *R*: bond distance; $\sigma^2$: Debye-Waller factors; Δ*E*: the inner potential correction. *R* factor: goodness of fit. * fitting with fixed parameter.

## 6. Recovery dynamics analysis of LN single crystal thin films *via* femtosecond transient absorption spectroscopy (fs-TAS).

fs-TAS profile and recovery dynamics analysis of lithium niobate (LN) single crystal thin films provide valuable insights into their optical and electronic properties, as well as their ultrafast relaxation processes. By employing fs-TAS, the temporal evolution of the excited states and photoinduced dynamics in LN single crystal thin films can be studied with high temporal resolution. The fs-TAS profile captures the transient absorption spectra as a function of time, revealing the dynamics of electronic excitations, charge carriers, and structural changes occurring on femtosecond to picosecond timescales. The analysis of fs-TAS data allows for the identification and characterization of various photogenerated species, such as excitons, polarons, and trapped charge carriers, as well as their respective lifetimes and relaxation pathways. This information is crucial for understanding the underlying mechanisms governing the optical and electronic behavior of LN thin films. Moreover, the recovery dynamics analysis provides insights into the recombination processes, charge carrier trapping, and relaxation mechanisms in LN single crystal thin films. By monitoring the time-dependent recovery of the transient absorption signal, one can extract important parameters such as carrier lifetimes, diffusion coefficients, and trap densities, which directly impact the material's performance in applications such as photovoltaics, optoelectronics, and nonlinear optics. The combination of fs-TAS profile and recovery dynamics analysis enables a comprehensive understanding of the ultrafast photophysical and photochemical processes occurring in LN single crystal thin films. This knowledge is fundamental for optimizing device performance, designing novel materials, and exploring new avenues for their applications in next-generation photonic and electronic technologies.

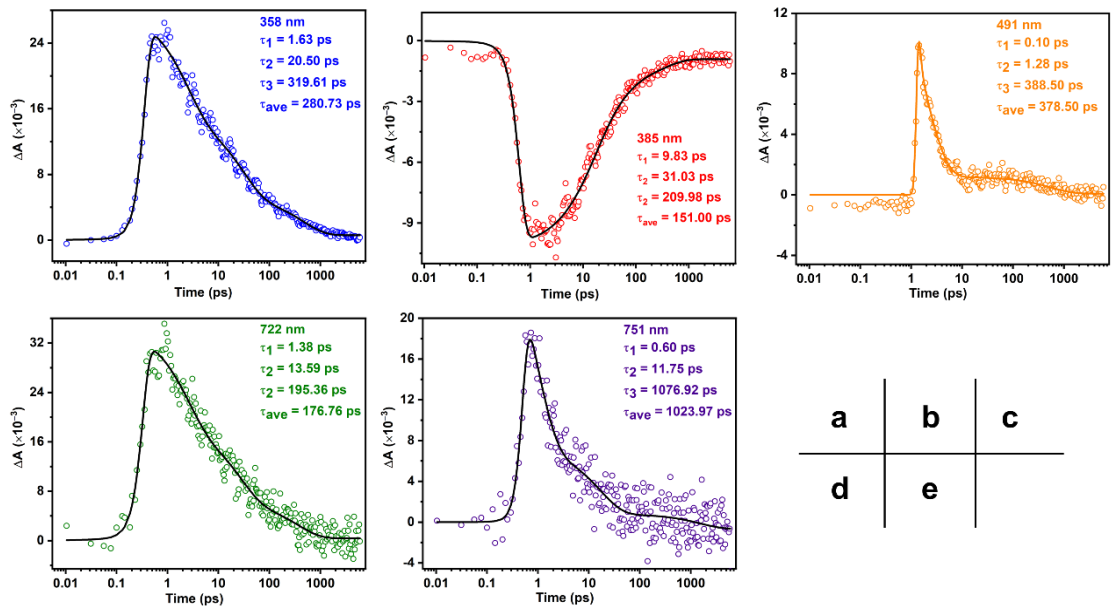

Supplementary Fig. 5| Normalized ΔA intensity as a function of delay time probed (c)358 nm, (d) 722 nm, (e) 385 nm, and (f) 751 nm.

## 7. Femtosecond transient absorption spectroscopy (fs-TAS).

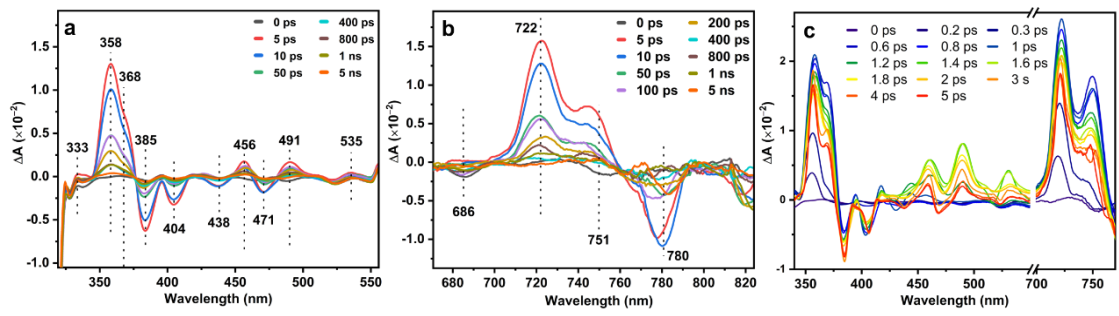

Supplementary Fig. 5| TA spectra recorded from LN single crystal thin film.

## 8. Electron-phonon oscillation analysis of LN single crystal thin films *via* femtosecond transient absorption spectroscopy (fs-TAS).

Femtosecond transient absorption spectroscopy (fs-TAS) is a powerful experimental method that involves exciting a material with an ultrafast laser pulse and probing its absorption spectrum with a time-delayed probe pulse. By carefully controlling the time delay between the excitation and probe pulses, researchers can monitor the rapid energy transfer processes occurring in the material on a femtosecond ($10^{15}$ seconds) timescale. In the case of LN single crystal thin films, the electron-phonon oscillation analysis using fs-TAS allows researchers to investigate the interactions between electrons and phonons, which are quantized lattice vibrations. When the material is excited with an intense laser pulse, electrons absorb energy and transition to higher energy states, leading to the generation of phonons. These phonons subsequently affect the electron dynamics through scattering processes. By analyzing the transient absorption spectra obtained from fs-TAS measurements, researchers can extract valuable information about the electron-phonon coupling strength, relaxation timescales, and energy transfer mechanisms in LN single crystal thin films. This analysis helps in understanding the fundamental processes governing the material's electronic and thermal properties, and it can guide the development of advanced electronic and optoelectronic devices based on LN.

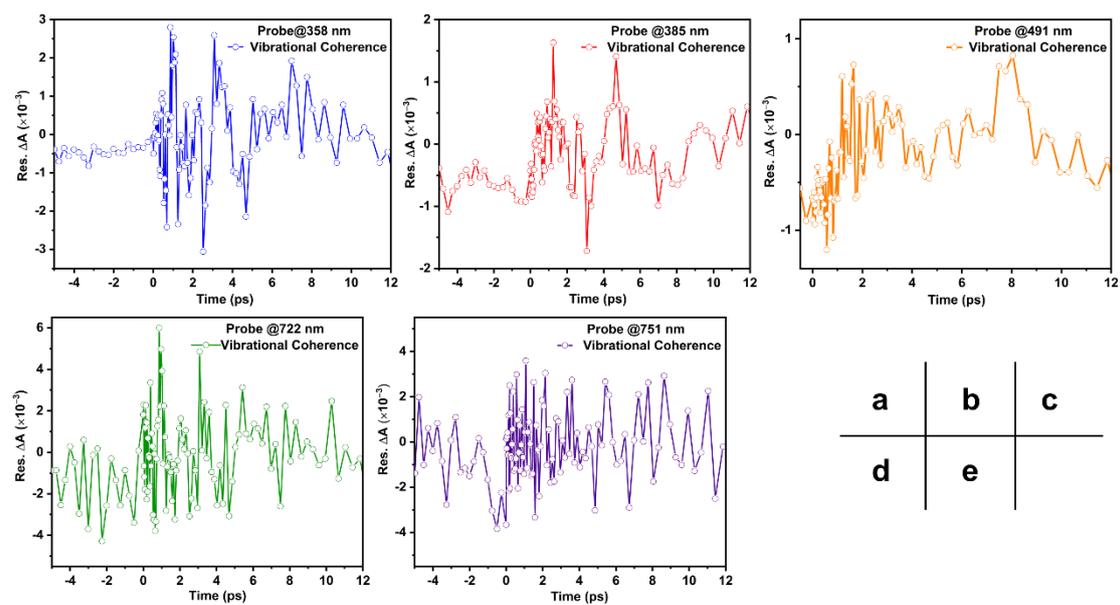

Supplementary Fig. 7| The extracted oscillatory parts of the pump–probe traces.

## 9. fs-TAS profile and .recovery dynamics analysis of LN single crystal thin films.

fs-TAS profile and recovery dynamics analysis of LN single crystal thin films offer a powerful approach to investigate the ultrafast optical properties and dynamic processes in these materials. With fs-TAS, the temporal evolution of the optical response in LN single crystal thin films can be probed with exceptional time resolution. The fs-TAS profile captures the changes in absorption as a function of time, providing valuable information about the excited state dynamics, carrier relaxation processes, and energy transfer pathways on femtosecond and picosecond timescales. By analyzing the fs-TAS data, it is possible to extract key parameters such as excited state lifetimes, relaxation rates, and energy transfer efficiencies. These parameters shed light on the underlying electronic and optical interactions in LN single crystal thin films and provide insights into the material's potential for applications in areas such as photonics, optoelectronics, and energy conversion. Furthermore, the recovery dynamics analysis focuses on understanding the relaxation and recombination processes in LN single crystal thin films following photoexcitation. By monitoring the time-dependent recovery of the transient absorption signal, valuable information about carrier lifetimes, diffusion coefficients, and recombination mechanisms can be obtained. The recovery dynamics analysis not only reveals the timescales involved in carrier relaxation and recombination but also provides insights into the presence of trap states, surface recombination, and other nonradiative processes. This information is crucial for optimizing the performance of LN single crystal thin films in devices such as solar cells, photodetectors, and nonlinear optical devices. In summary, the combined use of fs-TAS profile and recovery dynamics analysis enables a comprehensive understanding of the ultrafast optical properties, carrier dynamics, and energy transfer processes in LN single crystal thin films. This knowledge serves as a foundation for advancing the design and optimization of LN-based devices and paves the way for future developments in the field of ultrafast optoelectronics.

Supplementary Table 2. Summary of lifetimes

| Entry | Wavelength | Rise time (ps) | Peak time (ps) | On set time (ps) |
|---|---|---|---|---|
| 1 | @358 nm | 0.253 | 0.557 | 0 |
| 2 | @385 nm | 0.486 | 0.991 | 0.434 |
| 3 | @491 nm | 0.236 | 1.439 | 0.882 |
| 4 | @722 nm | 0.258 | 0.557 | 0 |
| 5 | @751 nm | 0.320 | 0.696 | 0.139 |

**10. Comparative analysis of bond lengths after structural relaxation in the LN crystal.**

The comparative analysis of bond lengths after structural relaxation in the LN crystal provides valuable insights into the atomic arrangements and bonding characteristics of this material. During structural relaxation, the LN crystal undergoes atomic movements and adjustments to reach a more stable configuration. This process leads to changes in the bond lengths, which are crucial for understanding the structural transformations and their implications on the material's properties. By comparing the bond lengths before and after structural relaxation, one can observe any deviations from the initial crystal structure and assess the level of atomic rearrangement. These comparative analyses help in identifying the regions of the crystal that are more prone to distortion or undergo significant changes in bonding. Furthermore, the analysis of bond lengths can provide information about the strength and nature of the chemical bonds in the LN crystal. Variations in bond lengths may indicate changes in the bond strength or the presence of different bonding environments, such as the formation of defects, substitutional impurities, or interstitial species. The comparative analysis of bond lengths can be complemented with other structural characterization techniques, such as X-ray diffraction or density functional theory calculations, to obtain a comprehensive understanding of the crystal's atomic arrangement and its impact on the material's properties. Overall, the comparative analysis of bond lengths after structural relaxation in the LN crystal allows for a detailed investigation of the structural changes and bonding characteristics. This knowledge contributes to the fundamental understanding of LN's properties and aids in the design and optimization of LN-based devices for various applications, including photonics, optoelectronics, and information storage.

Supplementary Table 3. Comparison of bond lengths between LN and LN-Nb$_{Li}$.

| Atom 1 | Atom 2 | d Å | |
|---|---|---|---|
| | | LN | LN-Nb$_{Li}$ |
| Li$_1$ | O$_{11}$ | 2.0123 | 1.9874 |
| | O$_{10}$ | 2.0125 | 1.993 |
| | O$_{12}$ | 2.0126 | 2.0025 |
| | O$_3$ | 2.3155 | 2.3118 |
| | O$_1$ | 2.3155 | 2.3147 |
| | O$_2$ | 2.3156 | 2.3381 |
| Li$_2$/Nb$_{Li}$ | O$_8$ | 2.0124 | 2.0333 |
| | O$_9$ | 2.0124 | 2.1024 |
| | O$_7$ | 2.0125 | 2.1029 |
| | O$_6$ | 2.3161 | 2.1242 |
| | O$_4$ | 2.3161 | 2.1301 |
| | O$_5$ | 2.3164 | 2.3084 |
| Li$_3$ | O$_{16}$ | 2.0125 | 2.0054 |
| | O$_{18}$ | 2.0126 | 2.0232 |
| | O$_{17}$ | 2.0126 | 2.0237 |
| | O$_8$ | 2.3162 | 2.2926 |
| | O$_7$ | 2.3162 | 2.2965 |
| | O$_9$ | 2.3163 | 2.3019 |
| Li$_4$ | O$_{14}$ | 2.0126 | 2.003 |
| | O$_{13}$ | 2.0129 | 2.0303 |
| | O$_{15}$ | 2.013 | 2.0378 |
| | O$_{10}$ | 2.3155 | 2.2327 |
| | O$_{12}$ | 2.3157 | 2.2434 |
| | O$_{11}$ | 2.3158 | 2.4175 |
| Li$_5$ | O$_5$ | 2.0121 | 1.9394 |
| | O$_4$ | 2.0122 | 1.9714 |
| | O$_6$ | 2.0124 | 2.0042 |
| | O$_{13}$ | 2.3161 | 2.116 |
| | O$_{15}$ | 2.3163 | |
| | O$_{14}$ | 2.3163 | |
| Li$_6$ | O$_1$ | 2.0125 | 1.9526 |
| | O$_2$ | 2.0125 | 1.9803 |
| | O$_3$ | 2.0127 | 2.0719 |
| | O$_{16}$ | 2.3161 | 2.2897 |
|     | O$_{18}$ | 2.3163 | 2.3293 |
|     | O$_{17}$ | 2.3163 | 2.4319 |
| Nb$_1$ | O$_{13}$ | 1.9514 | 1.9303 |
|     | O$_{14}$ | 1.9514 | 1.9488 |
|     | O$_{15}$ | 1.9515 | 1.9528 |
|     | O$_6$ | 2.1114 | 2.1123 |
|     | O$_4$ | 2.1115 | 2.1397 |
|     | O$_5$ | 2.1118 | 2.1961 |
| Nb$_2$ | O$_{16}$ | 1.9518 | 1.9456 |
|     | O$_{17}$ | 1.9518 | 1.9463 |
|     | O$_{18}$ | 1.9519 | 1.9599 |
|     | O$_3$ | 2.1114 | 2.0914 |
|     | O$_1$ | 2.1114 | 2.1085 |
|     | O$_2$ | 2.1115 | 2.1315 |
| Nb$_3$ | O$_3$ | 1.9515 | 1.9424 |
|     | O$_2$ | 1.9515 | 1.9537 |
|     | O$_1$ | 1.9516 | 1.957 |
|     | O$_{11}$ | 2.1115 | 2.0952 |
|     | O$_{10}$ | 2.1116 | 2.1145 |
|     | O$_{12}$ | 2.1118 | 2.1186 |
| Nb$_4$ | O$_5$ | 1.9518 | 1.9371 |
|     | O$_6$ | 1.9519 | 1.9397 |
|     | O$_4$ | 1.952 | 1.9851 |
|     | O$_8$ | 2.1115 | 2.0906 |
|     | O$_7$ | 2.1115 | 2.1172 |
|     | O$_9$ | 2.1116 | 2.1228 |
| Nb$_5$ | O$_7$ | 1.9519 | 1.9989 |
|     | O$_9$ | 1.9519 | 2.035 |
|     | O$_8$ | 1.9521 | 2.0451 |
|     | O$_{16}$ | 2.1115 | 2.064 |
|     | O$_{18}$ | 2.1116 | 2.137 |
|     | O$_{17}$ | 2.1117 | 2.1479 |
| Nb$_6$ | O$_{12}$ | 1.9516 | 2.0606 |
|     | O$_{10}$ | 1.9519 | 2.0718 |
|     | O$_{11}$ | 1.9521 | 2.1127 |
|     | O$_{13}$ | 2.1114 | 2.157 |
|     | O$_{15}$ | 2.1116 | 2.1881 |
|     | O$_{14}$ | 2.1117 | 2.2496 |

## 11. Band structure analysis based on the LN model.

Analysis of the band structure of LN reveals important insights into its electronic properties. The band structure of LN describes the distribution of electronic energy levels and momentum in the material. The band structure analysis of LN based on the LN model provides valuable information about its electronic behavior. It helps determine the bandgap, dispersion relations, and density of states, which are crucial for understanding the material's conductivity, insulating properties, or semiconductor characteristics. By solving the equations of the LN model, we can obtain the bandgap and examine the band crossings, overlaps, and couplings, which are essential for understanding the material's electronic transport and optoelectronic properties. The analysis of the band structure of LN using the LN model is particularly useful in material design, optimizing semiconductor devices, and studying optoelectronic devices. Theoretical calculations and experimental measurements provide insights into the band structure and further investigations of the electronic behavior, light absorption, and emission properties of LN. In summary, the analysis of the band structure of LN based on the LN model is a significant research method that allows a deeper understanding of its electronic energy levels and characteristics. It provides valuable guidance for materials science and device design.

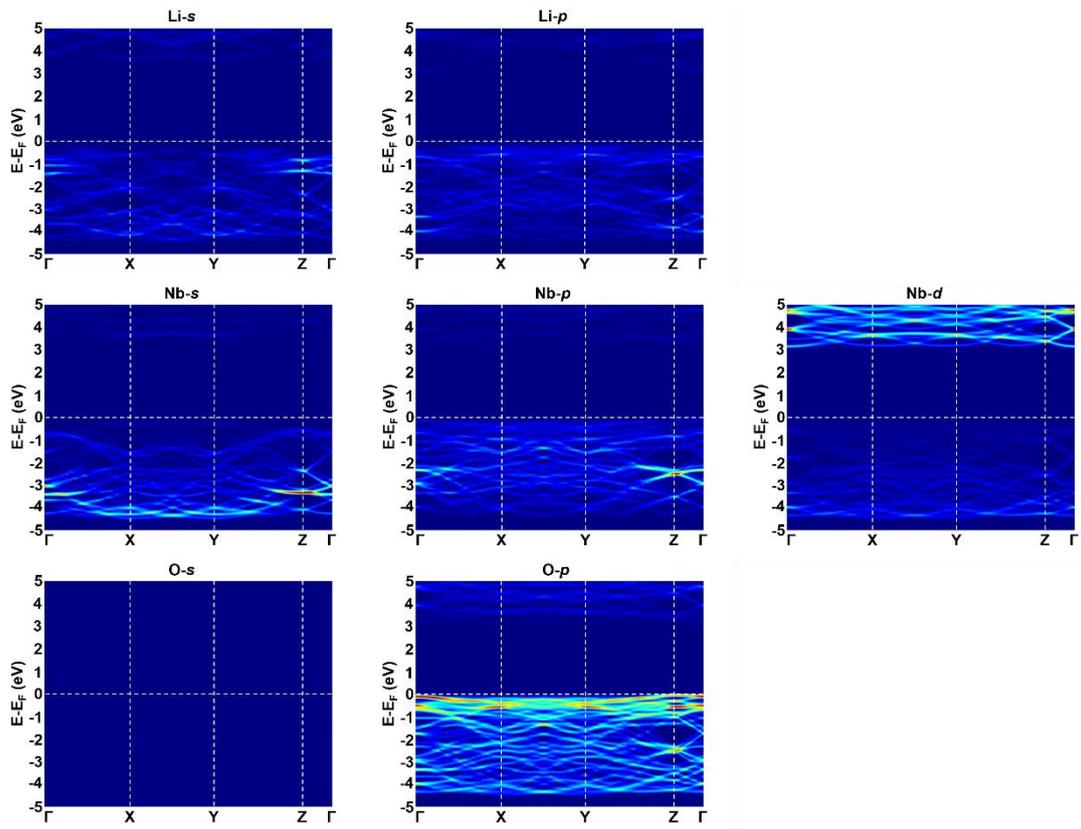

Supplementary Fig. 8| Orbital-projected band structure based on the LN model.

## 12. Band structure analysis based on the LN-Nb$_{Li}$ model.

Band structure analysis based on the LN-Nb$_{Li}$ model provides valuable insights into the electronic properties of LN with Nb$_{Li}$ antisite defects. The presence of NbLi defects introduces structural and electronic perturbations in the crystal lattice, which significantly impact the material's band structure. The LN-Nb$_{Li}$ model takes into account the interactions and effects caused by the presence of Nb$_{Li}$ anti-site defects in LN. By considering the altered crystal structure and electronic configuration resulting from these defects, the LN-Nb$_{Li}$ model allows us to analyze the band structure of LN with high precision. The band structure analysis reveals the changes in the electronic energy levels and momentum distribution due to the Nb$_{Li}$ defects. It provides crucial information about the defect-induced bandgap modifications, dispersion relations, and density of states. The presence of Nb$_{Li}$ defects can lead to localized states within the bandgap, resulting in the creation of defect levels or mid-gap states. These states can significantly influence the material's electrical conductivity and optical properties. The band structure analysis based on the LN-Nb$_{Li}$ model helps in characterizing the nature and spatial distribution of these defect states. Understanding the band structure of LN with Nb$_{Li}$ defects is of great importance in various applications. It aids in the design and optimization of LN-based devices, such as waveguides, modulators, and sensors, where the presence of defects can affect the performance and functionality of the devices. In summary, the band structure analysis based on the LN-Nb$_{Li}$ model provides a detailed understanding of the electronic behavior of LN with Nb$_{Li}$ antisite defects. It enables us to comprehend the impact of these defects on the material's bandgap, electronic states, and optical properties, thereby guiding the development of defect-engineered LN materials and devices.

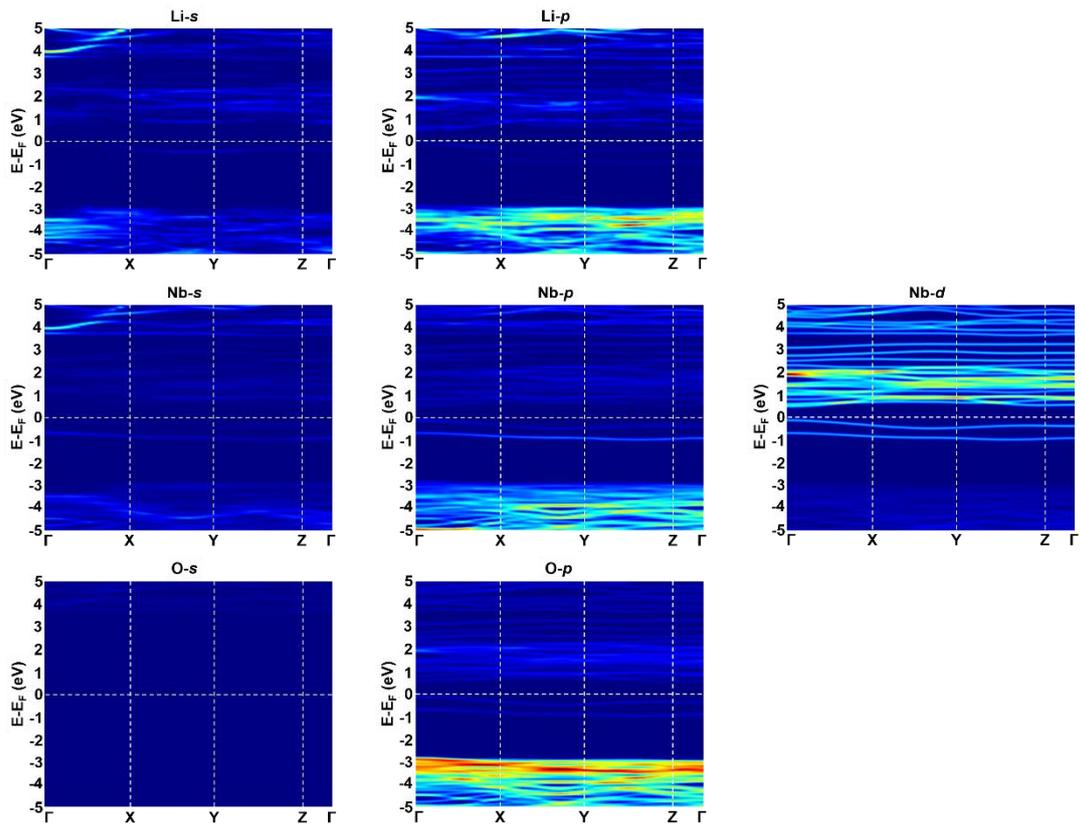

Supplementary Fig. 9| Orbital-projected band structure based on the LN-Nb$_{Li}$ model.